\documentclass[a4paper,fleqn,usenatbib]{mnras}

\usepackage[T1]{fontenc}
\usepackage{ae,aecompl}

\usepackage{graphicx}	
\usepackage{amsmath}	
\usepackage{amssymb}	
\usepackage{cite}
\usepackage{subcaption}
\usepackage{float}
\restylefloat{table}
\restylefloat{figure}

\title[Role of galactic bars in the formation of spiral arms]{Role of galactic bars in the formation of spiral arms: A study through orbital and escape dynamics - I}

\author[Mondal \& Chattopadhyay]{Debasish Mondal$^{1}$\thanks{E-mail: dmappmath$\_$rs@caluniv.ac.in}
and Tanuka Chattopadhyay$^{1}$\thanks{E-mail: tchatappmath@caluniv.ac.in}\\
$^{1}$ Department of Applied Mathematics, University of Calcutta, $92$ A. P. C. Road, Kolkata $700009$, India}

\date{Accepted XXX. Received YYY; in original form ZZZ}

\pubyear{2020}

\begin{document}
\label{firstpage}
\pagerange{\pageref{firstpage}--\pageref{lastpage}}
\maketitle

\begin{abstract}
In the present work we have developed a three-dimensional gravitational model of barred galaxies, in order to study orbital and escape dynamics of the stars inside their central barred region. Our gravitational model is composed of four components, central nucleus, bar, disc and dark matter halo. Furthermore we have analysed the model for two different types of bar potentials. The study has been carried out for a Hamiltonian system and thorough numerical studies have been done in order to categorize regular and chaotic motions of stars. We have seen that escape mechanism has only seen near saddle points ($L_2$, $L_4$ and $L_2^{'}$, $L_4^{'}$) of the Hamiltonian system. Orbital structures in $x$ - $y$ plane indicate that this escaping motion corresponds to the two ends of the bar. Classifications of orbits are found by calculating maximal Lyapunov exponent of the stellar trajectories corresponding to a specific initial condition vector. Poincaré surface section maps are studied in both $x$ - $y$ and $x$ - $p_x$ ($p_x$ is the momentum along $x$ - direction) plane to get a complete view of the escape properties of the system in the phase space. Also we studied in detail how the chaotic dynamics varies with mass, length and nature of the bar. We found that under suitable physical conditions the chaos plays a pivotal role behind the formation of grand design or poor spiral pattern for stronger bars and ring structures for weaker bars.   
\end{abstract}

\begin{keywords}
Galaxy: kinematics and dynamics -- galaxies: structure -- galaxies: bar -- chaos
\end{keywords}



\section{Introduction}
\label{sec:1}
In Hubble's classification of galaxies central stellar bar structure is mainly observed in lenticular (example: NGC 1460, NGC 1533, NGC 2787) and spiral galaxies (example: M58, M91, M95, M109, NGC 1300, NGC 1365, NGC 1512, NGC 2217, NGC 2903, NGC 3953, NGC 4314, NGC 4921, NGC 7541, UGC 12158). Also some irregular galaxies like Large Magellanic Cloud (LMC) and Small Magellanic Cloud (SMC) have off-centred stellar bar \citep{Zhao2000, Bekki2009, Piatti2017, Monteagudo2018, Strantzalis2019}, though not all lenticular and spiral galaxies have stellar bars. All lenticulars and spirals are disc supported systems and these stellar discs may or may not support stellar bars. Only one third of the local disc galaxies have determinable type of bars (or strong bars) and another one third have indeterminable type of bars (or weak bars) \citep{Eskridge2000, Cheung2013, Yoon2019}. Fraction of barred galaxies among lenticulars and spirals is strongly dependent on red-shift, stellar mass, colour and bulge prominence \citep{Abraham1999, Sheth2008, Nair2010, Simmons2014, Zhou2020}.

\indent Galactic bars are one of the robust substructure of the barred galaxies. They are solid, dense stellar body rotating around the central core. Pattern speed of the bar is different than that of the disc. There are many theories behind origin of the galactic bar \citep{Miwa1998, Bournaud2002, Seo2019, Petersen2019, Polyachenko2020}. Most evident theory is galactic bars are rotational instabilities, which arises due to density waves radiating outwards from the galactic core. These instabilities influence stellar orbits by redistributing their trajectories. As time goes these reshaped orbits follow an outward motion, which further creates a self stabilizing stellar structure, in the form of bar \citep{Raha1991, Sellwood2016, Bovy2019, Lokas2019, Sanders2019}. Barred galaxies mostly have single bar structure embedded inside the bulge, though there are many examples of double barred galaxies also (example: Milky Way, NGC 1291, NGC 1326, NGC 1543). In such cases the smaller secondary bar is wrapped inside the larger primary bar \citep{Erwin2004, Debattista2006, Du2016, Lorenzo2019}.

\indent Galactic bars have many shapes and sizes according to the functional form of potential energy. There are many three-dimensional bar potential models like spherical, homeoidal, ellipsoidal etc. \citep{Ferrers1877, Vaucouleurs1972, Caranicolas2002, Jung2015, Williams2017} have extensively studied till date. Ferrers' triaxial potential \citep{Ferrers1877} is the most used realistic bar potential model, though its functional form is very much complex and also it is computationally very much challenging. There are some other potentials like homeoidal potential \citep{Vaucouleurs1972}, ad hoc potential \citep{Barbanis1967, Dehnen2000} etc. have simpler functional forms than Ferrers' but still they are rigorous to handle numerically. Also there are some simple realistic bar potential models used by \citet{Caranicolas2002}, \citet{Jung2015} etc.

\indent Due to influence of Galactic bar some of the stellar orbits remain trapped inside the potential boundary and while others are escaped from that boundary, during their time evolution. This problem can be studied from viewpoint of the problem of escape in an open Hamiltonian dynamical system \citep{Contopoulos2004, Ernst2008, Ernst2014, Jung2015, Jung2016}. An open Hamiltonian system is a system where for energies above an escape threshold, the energy shell is non compact and as a result a part of the stellar orbits explores (here from potential holes to saddles) an infinite part of the position space. Also the Hamiltonian dynamics is a time reversal invariant \citep{Jung2016}. Now for a conservative dynamical system the Hamiltonian (or the total energy) is a constant of motion. Hence all the stellar orbits are confined inside the five-dimensional energy hyper-surface of the six-dimensional phase space of the Hamiltonian system. These stellar orbits are either regular or chaotic in nature according to their initial condition (or initial energy value). Orbits having initial energy below the escape energy are remain trapped inside the potential boundary and exhibit bounded motion (regular or chaotic). While orbits having initial energy above the escape energy exhibit unbounded motion. They escape from the interior along the open zero velocity curves (exist along the escape channels of the potential). For bounded motion there are orbits which initially look like regular orbits but revealed their true chaotic nature in long time period. These orbits are known as trapped chaotic orbits. Existence of such orbits make the stellar dynamics more geometrically complicated in the phase space. For unbounded motion orbits are generally chaotic in nature. There are many dynamical indicators which can classify these orbits according to their regular or chaotic behaviour. Lyapunov exponent is one such effective dynamical indicator and its working mechanism is quite simple \citep{Sandri1996}. It calculates the rate of separation of two neighbouring trajectories during the entire time period of evolution. Value of the Maximal Lyapunov Exponent (hereafter MLE) gives us more complete view about dynamical nature of these orbits (regular or chaotic) in the vast sea of initial conditions of the phase space. MLE basically is the highest separation between two neighbouring trajectories starting from same initial condition in a designated time interval. If the value of MLE is positive then it indicates that orbits are chaotic in nature, while MLE $=0$ indicates that orbits are periodic in nature \citep{Strogatz1994}. To analyse escape properties of these orbits, one need to visualise Poincaré surface section maps \citep{Birkhoff1927} in different two-dimensional phase planes. Under suitable physical conditions these escaped orbits through the potential saddles further fuel the formation of spiral arms, which means there is some kind of bar driven spiral arm formation mechanism in case of barred spiral galaxies. Effect of dynamical chaos of the stellar orbits behind formation of the spiral arms have extensively studied in the recent past \citep{Lindblad1947, Bell1972, Pfenniger1984, Patsis2012, Mestre2020}.   

\indent Most of the earlier studies focused on detecting the chaotic invariant set of stellar orbits, which governs the general dynamics in the central region. Also there are studies about how these chaotic invariant set fuels the formation of spiral arms. Due to chaotic dynamics in the central region, the escaping stars produce tidal trails at the two ends of the bar. These tidal trails have tendency to create different morphologies like ring or spiral arms due to non-axisymmetric perturbations. These studies relates the fate of escaping stars with subsequent morphological structures \citep{Di2005, Minchev2010, Quillen2011, Grand2012, Onghia2013, Ernst2014, Jung2016}. \\
\indent In the present work we showed the same analogy i.e. the fate of escaping stars behind formation of spiral arms but from the viewpoint of the amount of chaos produced therein. Therefore we primarily focus on detection of the chaotic dynamics in the central region of barred galaxies, under suitable realistic bar potential models. We then figure out a comparison between these bar potential models in order to show how these models affect the chaotic dynamics with respect to mass and length of the bar. We relate these measurements of chaos with subsequent structure formations like spiral arms or rings. Also we have shown which bar model is more feasible for certain type of structure formations under suitable astrophysical circumstances.

\indent Here we used a four component three-dimensional gravitational model for the barred galaxies. The model consists of a spherical bulge, a bar embedded in bulge, a flat disc and a logarithmic dark matter halo. Modelling is done in two parts corresponding to bar potential models, (i) an anharmonic mass-model bar potential  \citep{Caranicolas2002} and (ii) the Zotos bar potential \citep{Jung2015}. These two are the most non arduous potential forms studied till date. For each of these potential models, first we have studied several orbital structures corresponding to different initial condition vectors. Also we calculate values of MLE for each of these initial condition vector, which gives us complete dynamical nature of these orbits (regular or chaotic). Then we draw Poincaré surface section maps in both $x$ - $y$ and $x$ - $p_x$ subsection of the phase space. These two surface section maps are important in order to visualise the motion along the galactic plane which contains the disc. Also these two surface section maps are plotted for different energy values (energies higher than the escape energy of the saddles of our gravitational potential models), in order to get idea about escape mechanism of the system. Finally we described how the nature of orbits varies with the bar parameters e.g. mass and length by calculating the MLE for each of the two bar potential models.    

Our work is divided into four sections. Section \ref{sec:2} describes the mathematical part of the barred galaxy model. Section \ref{sec:2} has divided into two subsections. In subsection \ref{sec:2.1} we have discussed the model for an anharmonic bar potential and in subsection \ref{sec:2.2} we have discussed the same model for the Zotos bar potential. Section \ref{sec:3} consists of numerical analysis part. Section \ref{sec:3} has divided into three subsections. In subsection \ref{sec:3.1} several orbital structures are plotted in $x$ - $y$ plane. In subsection \ref{sec:3.2} several Poincaré surface section maps are plotted in both $x$ - $y$ and $x$ - $p_x$ planes while in subsection \ref{sec:3.3} we have discussed how the chaotic dynamics evolved with mass and length of the bar under the influence of the two bar potential models. Finally discussion and conclusions are given in Section \ref{sec:4}.

\section{Gravitational Theory}
\label{sec:2}
\subsection{Model 1}
\label{sec:2.1}
We consider a three-dimensional gravitational model of barred galaxies and investigate orbital motions of stars inside their central region. Our gravitational model has four components -- (i) spherical bulge, (ii) bar embedded in the bulge, (iii) flat disc and (iv) logarithmic dark matter halo. Here all the modelling and calculations are done in the Cartesian co-ordinate system. Let $\Phi_\text{t}(x,y,z)$ be the total potential of the galaxy. This $\Phi_\text{t}(x,y,z)$ consists of four parts and they are  (i) bulge potential -- $\Phi_\text{B}(x,y,z)$, (ii) bar potential -- $\Phi_\text{b}(x,y,z)$, (iii) disc potential -- $\Phi_\text{d}(x,y,z)$ and (iv) dark matter halo potential -- $\Phi_\text{h}(x,y,z)$. Therefore, 
$$\Phi_\text{t}(x,y,z) = \Phi_\text{B}(x,y,z) + \Phi_\text{b}(x,y,z) + \Phi_\text{d}(x,y,z) + \Phi_\text{h}(x,y,z).$$

\noindent Density distribution $\rho_\text{t}(x,y,z)$ corresponds to $\Phi_\text{t}(x,y,z)$ is given through Poisson equation,
$${\nabla}^2 \Phi_\text{t} (x,y,z) = 4 \upi G \rho_\text{t}(x,y,z),$$

\noindent where $G$ is the gravitational constant. Now let $\vec{\Omega_\text{b}} \equiv (0,0,\Omega_\text{b})$ be the constant angular velocity of the bar, which follows a clockwise rotation along $z$ - axis. In this rotating frame the effective potential is, 

\begin{equation}
\label{eq:1}
\Phi_{\text{eff}}(x,y,z) = \Phi_\text{t}(x,y,z) - \frac{1}{2} \Omega_\text{b}^2 (x^2 + y^2).
\end{equation} 

The potential functions for different sub-structures are as follows,
\begin{itemize}
\item Bulge: Central bulge is the excess of luminosity from the surrounding galactic disc. The distribution of stars in the galactic bulges are not exponential rather spherically symmetric and dominated by old red stars. That's why we choose a potential of the Plummer type \citep{Plummer1911} to describe the distribution of matter inside the bulge of barred galaxies \citep{Binney1987, Sofue2001, Halle2013, Salak2016}. Now density distribution of matter in the bulge for Plummer potential is,
$$ \rho_\text{B}(x,y,z) = \frac{3 M_\text{B} c_\text{B}^2}{4 \pi} \frac{1}{(x^2 + y^2 +z^2 + c_\text{B}^2)^\frac{5}{2}},$$
and associated form of the potential is,
$$
\Phi_\text{B}(x,y,z) = - \frac{G M_\text{B}}{\sqrt{x^2 + y^2 + z^2 + c_\text{B}^2}},
$$
where $M_\text{B}$ is the mass of the bulge and $c_\text{B}$ is the radial scale length. Here we consider a massive dense bulge rather than a central supermassive black hole, so that we can exclude all relativistic effects from our model.

\item Bar: Bar is an extended linear non-axisymmetric stellar structure in the central region of a galaxy. For model $1$, we choose a strong bar potential, whose density in the central region is very high (i.e. a cuspy type, see Fig. \ref{fig:3}). For this we consider an anharmonic mass-model potential \citep{Caranicolas2002}, which has a non arduous potential form and its density distribution of matter is,
$$\rho_\text{b}(x,y,z) = \frac{M_\text{b}}{4 \pi} \frac{[(2 + \alpha^2) c_{\text{b}}^2 - (1 - \alpha^2)(x^2 - 2 \alpha^2 y^2 + z^2)]}{{(x^2 + \alpha^2 y^2 + z^2 + c_{\text{b}}^2)}^\frac{5}{2}},$$
and associated form of the potential is,
$$
\Phi_\text{b}(x,y,z) = - \frac{G M_\text{b}}{\sqrt{x^2 + (\alpha y)^2 + z^2 + c_\text{b}^2}},
$$
where $M_\text{b}$ is the mass of the bar, $\alpha$ is the bar flattening parameter and $c_\text{b}$ is the radial scale length.

\item Disc: Disc is the most luminous component of a galaxy. Distribution of matter inside the disc is axisymmetric, flattened and exponentially falls off with galactocentric radius. Structure of the disc can be thought as a flattened spheroid. For this flattened spheroid disc structure \citep{Binney1987, Flynn1996, Shin2007, Smet2015, An2019}, we use the gravitational potential model developed by Miyamoto and Nagai \citep{Nagai1975}, often termed as Miyamoto and Nagai potential. This potential model has a simple analytical form for the corresponding density distribution of matter,
$$\rho_\text{d}(x,y,z) = \frac{M_\text{d} h^2}{4 \pi}$$ 
$$\times \frac{k (x^2 + y^2) + (k + 3 \sqrt{h^2 + z^2}) {(k + \sqrt{h^2 + z^2})}^2}{{[(x^2 + y^2) + {(k + \sqrt{h^2 +z^2})}^2]}^{\frac{5}{2}} {(h^2 +z^2)}^{\frac{3}{2}}}.$$
Also associated form of the potential is,
$$
\Phi_\text{d}(x,y,z) = - \frac{G M_\text{d}}{\sqrt{x^2 + y^2 + (k + \sqrt{h^2 + z^2})^2}},
$$
where $M_\text{d}$ is the mass of the disc and $k$, $h$ are the horizontal and vertical scale lengths respectively.

\item Dark matter halo: Dark matter halo is the extended distribution of non-luminous (non-baryonic) matter of a galaxy \citep{Ostriker1974}. Structure of the dark matter haloes in barred galaxies are flattened axisymmetric, as observed rotation curve becomes almost flat at larger galactocentric distances. For this flattened axisymmetric dark matter halo structure \citep{Binney1987, Ernst2014}, we use a variant of logarithmic potential \citep{Zotos2012}. The corresponding density distribution of matter is,
$$ \rho_\text{h}(x,y,z) = \frac{v_{0}^2}{4 \pi G} \frac{\beta^2 x^2 + (2 \beta^2 - \beta^4) y^2 + \beta^2 z^2 + (\beta^2 + 2) c_\text{h}^2}{{(x^2 + \beta^2 y^2 +z^2 + c_\text{h}^2)}^2}$$
and associated form of the potential is,
$$
\Phi_\text{h}(x,y,z) = \frac{v_0^2}{2} \; \ln(x^2 + \beta^2 y^2 + z^2 + c_\text{h}^2),
$$
where $v_0$ is the circular velocity of halo, $\beta$ is the halo flattening parameter and $c_\text{h}$ is the radial scale length.	
\end{itemize}

\noindent In this model we take the physical units as --
\begin{itemize}
\item Unit of length: $1$ kpc
\item Unit of mass: $2.325 \times 10^7 M_\odot$
\item Unit of time: $0.9778 \times 10^8$ yr
\item Unit of velocity: $10$ km $\text{s}^{-1}$
\item Unit of angular momentum: $10$ km $\text{kpc}$ $\text{s}^{-1}$ 
\item Unit of energy per unit mass: $100$ $\text{km}^2$ $\text{s}^{-2}$.
\end{itemize} 
Without loss of any generality, we consider $G = 1$. The values of other physical parameters are taken from \citet{Zotos2012} and \citet{Jung2016} and are given in Table \ref{tab:1}. 

\begin{table}[H]
\centering
\caption{Model $1$ -- Values of physical parameters.}	
\label{tab:1}
\begin{tabular}{|c|c|}
\hline
Parameter    & Value\\
\hline
\hline
$M_\text{B}$ & 400\\
$c_\text{B}$ & 0.25\\
$M_\text{b}$ & 3500\\
$c_\text{b}$ & 1\\
$\alpha$     & 2\\
$M_\text{d}$ & 7000\\
\hline
\end{tabular}
\begin{tabular}{|c|c|}
\hline
Parameter         & Value\\
\hline
\hline
$k$               & 3\\
$h$               & 0.175\\
$v_0$             & 15\\
$\beta$           & 1.3\\
$c_\text{h}$      & 20\\
$\Omega_\text{b}$ & 1.25\\
\hline
\end{tabular}
\end{table}

\begin{figure}[H]
\centering
\includegraphics[height=8cm,width=\linewidth]{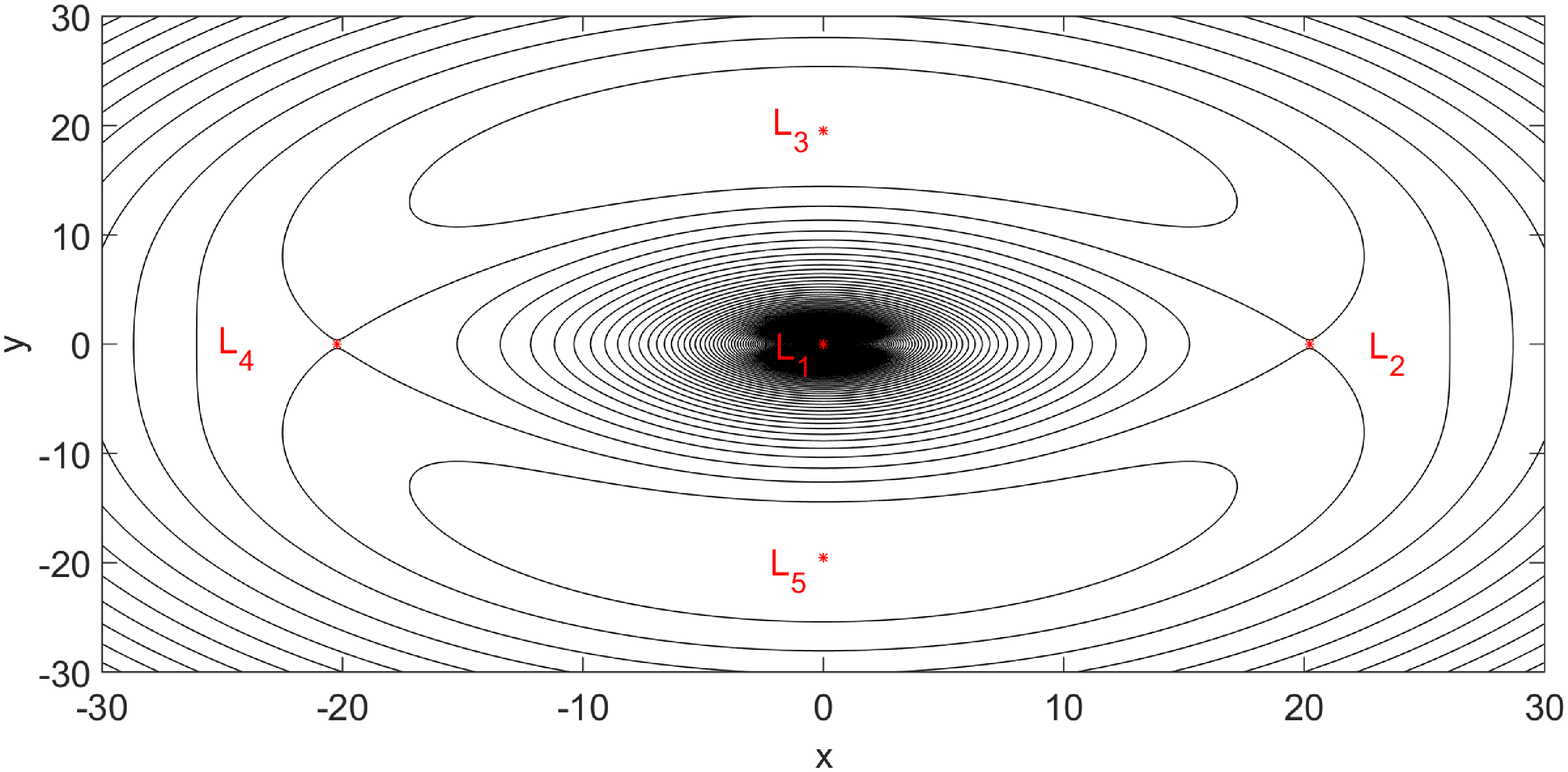}
\caption{Model $1$ -- The isoline contours of $\Phi_\text{eff}(x,y,z)$ in $x$ - $y$ plane at $z = 0$. Location of the Lagrangian points are marked in red.}
\label{fig:1}
\end{figure}

Now for a test particle (star) of unit mass the Hamiltonian ($H$) of the given system is defined as,
\begin{equation}
\label{eq:2}
H = \frac{1}{2} (p_x^2 + p_y^2 + p_z^2) \; + \; \Phi_\text{t}(x,y,z) \; - \;  \Omega_\text{b} L_z = E,
\end{equation}
where $\vec{r} \equiv (x,y,z)$ is the position vector of test particle at time $t$, $\vec{p} \equiv (p_x,p_y,p_z)$ is the corresponding linear momentum vector, $E$ is the total energy and $L_z = x p_y - y p_x$ is the $z$-component of the angular momentum vector, $\vec{L} = \vec{r} \times \vec{p}$. Hence Hamilton's equations of motion are, 

\begin{equation}
\label{eq:3}
\begin{split}
\dot{x} = p_x + \Omega_\text{b} y,\\
\dot{y} = p_y - \Omega_\text{b} x,\\
\dot{z} = p_z ,\\
\dot{p_x} = - \frac{\partial \Phi_\text{t}}{\partial x} + \Omega_\text{b} p_y,\\
\dot{p_y} = - \frac{\partial \Phi_\text{t}}{\partial y} - \Omega_\text{b} p_x,\\
\dot{p_z} = - \frac{\partial \Phi_\text{t}}{\partial z},\\
\end{split}
\end{equation}

\noindent where $`\cdot$' $\equiv \frac{d}{dt}$. This autonomous Hamiltonian system has five Lagrangian (or equilibrium) points namely $L_1$, $L_2$, $L_3$, $L_4$ and $L_5$, which are solutions of the following equations:
\begin{equation}
\label{eq:4}
\frac{\partial \Phi_\text{eff}}{\partial x} = 0, \;\;\; 
\frac{\partial \Phi_\text{eff}}{\partial y} = 0, \;\;\; 
\frac{\partial \Phi_\text{eff}}{\partial z} = 0.
\end{equation}
\noindent Locations of these Lagrangian points and their stability nature are given in Fig. \ref{fig:1} and Table \ref{tab:2} respectively. 

\begin{table}[H]	
\centering
\begin{tabular}{|c|c|c|}
Name  & Location             & Local Stability Type\\
\hline
\hline
$L_1$ & $(0,0,0)$            & Stable Centre\\
$L_2$ & $(20.23113677,0,0)$  & Unstable Saddle Point\\
$L_3$ & $(0,19.49818983,0)$  & Asymptotically Stable Node\\
$L_4$ & $(-20.23113677,0,0)$ & Unstable Saddle Point\\
$L_5$ & $(0,-19.49818983,0)$ & Asymptotically Stable Node\\
\hline
\end{tabular}
\caption{Model $1$ -- Stability nature of Lagrangian points.}
\label{tab:2}
\end{table}

\noindent The values of $E$ (or Jacobi value of integral of motion) at $L_2$ and $L_4$ are identical and that value is $E_{L_2} = -100.82313737 = E_{L_4}$. Similarly the values of $E$ at $L_3$ and $L_5$ are identical and that value is $E_{L_3} = 20.21395566 = E_{L_5}$. Also the value of E at $L_1$ is $E_{L_1} = -6630.68464790$. The nature of orbits in different energy range are given as --
\begin{itemize}
\item $E_{L_1} \leq E < E_{L_2}$: In this energy range motion of orbits are bounded.
\item $E \geq E_{L_2}$: In this energy range motion of orbits are unbounded and two symmetrical escape channels exist near both ${L_2}$ and ${L_4}$.
\end{itemize}

\subsection{Model 2} 
\label{sec:2.2}
\begin{figure}[H]
\centering
\includegraphics[height=8cm,width=\linewidth]{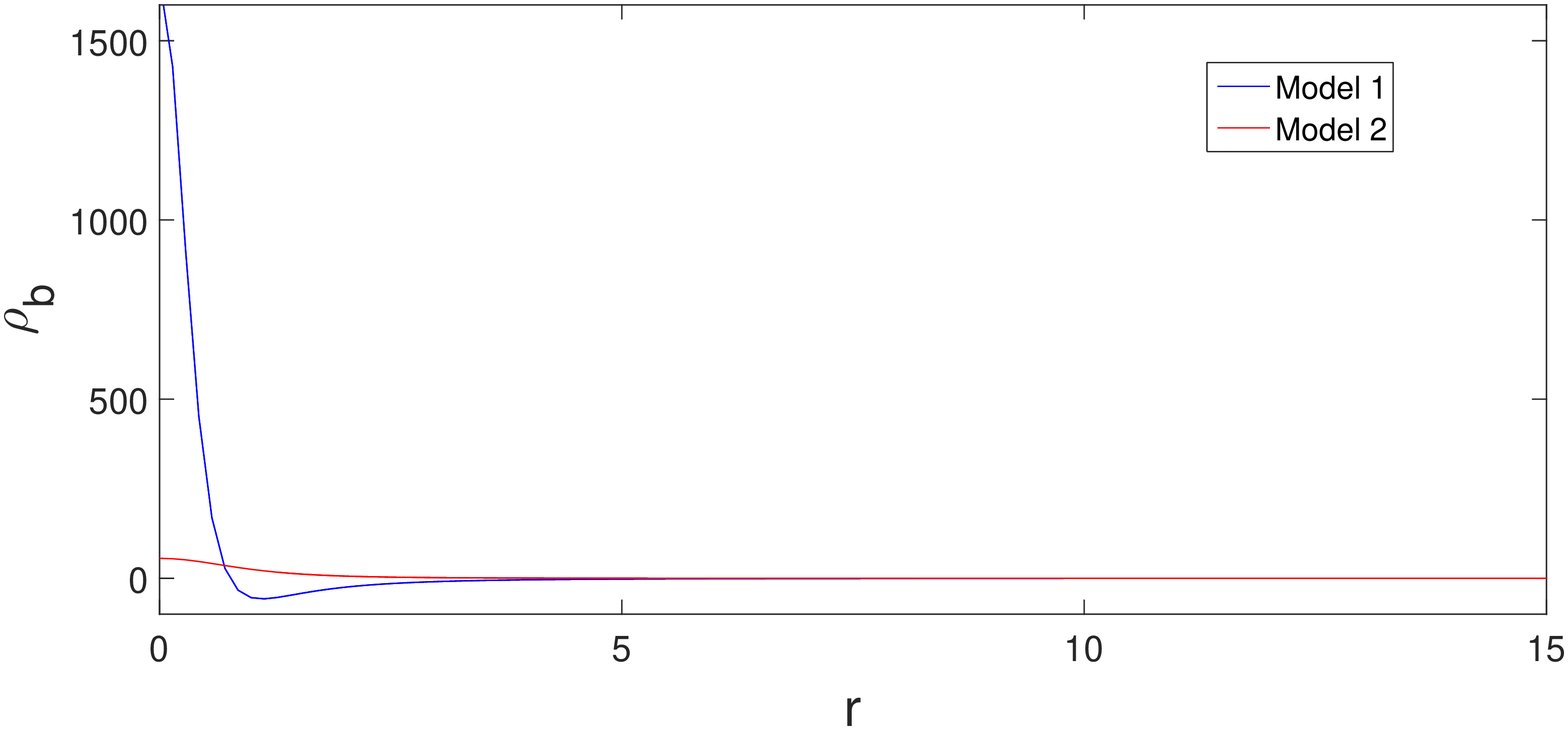}
\caption{Radial ($r = \sqrt{x^2 + y^2}$) distribution of the density structure of the bar ($\rho_\text{b}$) for both models $1$ and $2$ at $z = 0$.}
\label{fig:3}
\end{figure}

In this model we consider the three-dimensional gravitational model as discussed in subsection \ref{sec:2.1}, but change only the potential form of the galactic bar. Here we choose a comparatively weak bar potential, whose density at the central region is moderate (Fig. \ref{fig:3}). For this model we used the bar potential model developed by \citet{Jung2015}, which has also a non arduous potential form. This model is often termed as Zotos bar potential. For this potential model density distribution of matter is,
$$\rho_\text{b}(x,y,z) = \frac{M_\text{b} c_\text{b}^2}{8\pi a} [f(x+a,y,z) - f(x-a,y,z)],$$
$$\text{where} \;\; f(x,y,z) = \frac{x (2x^2 + 3y^2 + 3z^2 + 3 {c_\text{b}^2})}{{(y^2 + z^2 + {c_\text{b}^2})}^2 {(x^2 + y^2 + z^2 + {c_\text{b}^2})}^\frac{3}{2}}$$ 
and associated form of the potential is,
$$\Phi_\text{b}(x,y,z) = \frac{G M_\text{b}}{2 a} \; \ln(\frac{x - a + \sqrt{{(x - a)}^2 + y^2 + z^2 + {c_\text{b}^2}}}{x + a + \sqrt{{(x + a)}^2 + y^2 + z^2 + {c_\text{b}^2}}}),$$
where $M_\text{b}$ is the mass of the bar, $a$ is the length of semi-major axis of the bar and $c_\text{b}$ is the radial scale length. For this bar potential we use $a = 10$ and the values of other all model parameters due to bulge, disc and dark matter halo remain the same as given in Table \ref{tab:1}. 

\begin{figure}[H]
\centering
\includegraphics[height=8cm,width=\linewidth]{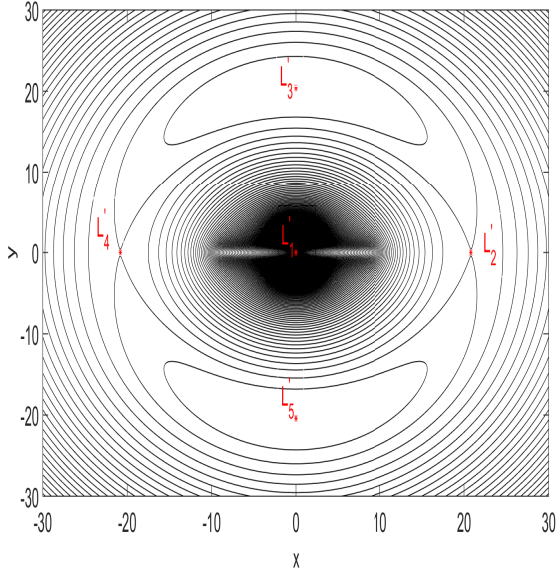}
\caption{Model $2$ -- The isoline contours of $\Phi_\text{eff}(x,y,z)$ in $x$ - $y$ plane at $z = 0$. Location of the Lagrangian points are marked in red.}
\label{fig:2}
\end{figure}

\noindent Locations of the Lagrangian points, namely ${L_1}^{'}$, ${L_2}^{'}$, ${L_3}^{'}$, ${L_4}^{'}$ \& ${L_5}^{'}$ and their stability natures are given in Fig. \ref{fig:2} and Table \ref{tab:3} respectively.

\begin{table}[H]	
\centering
\begin{tabular}{|c|c|c|}
Name      & Location             & Local Stability Type\\
\hline
\hline
$L_1^{'}$ & $(0,0,0)$            & Stable Centre\\
$L_2^{'}$ & $(20.82978638,0,0)$  & Unstable Saddle Point\\
$L_3^{'}$ & $(0,20.36721028,0)$ & Asymptotically Stable Node\\
$L_4^{'}$ & $(-20.82978638,0,0)$ & Unstable Saddle Point\\
$L_5^{'}$ & $(0,-20.36721028,0)$ & Asymptotically Stable Node\\
\hline
\end{tabular}
\caption{Model $2$ -- Stability nature of Lagrangian points.}
\label{tab:3}
\end{table}

\noindent The values of $E$ at $L_2^{'}$ and $L_4^{'}$ are identical and that value is $E_{L_2^{'}} = -116.46144116 = E_{L_4^{'}}$. Similarly the values of $E$ at $L_3^{'}$ and $L_5^{'}$ are identical and that value is $E_{L_3^{'}} = -60.76593330 = E_{L_5^{'}}$. Also the value of $E$ at $L_1^{'}$ is $E_{L_1^{'}} = -4180.06268050$. The nature of orbits in different energy range are same as discussed in model $1$.

\section{Numerical Results}
\label{sec:3}
To study the orbital and escape dynamics of stars along the galactic plane (which contains the bar), we put $z = 0 = p_z$ in both the gravitational models $1$ and $2$ respectively. Depending upon the initial energy value, stellar orbits are either trapped or escaped from the system. We have seen that two symmetrical escape channels exist only near $L_2$, $L_4$ and $L_2^{'}$, $L_4^{'}$ (Figs. \ref{fig:1} and \ref{fig:2}). Hence escape mechanism is only relevant near those Lagrangian points. Studying nature of orbits near either of $L_2$ or $L_4$ or of $L_2^{'}$, $L_4^{'}$ gives us complete information about the escape dynamics of the system. In order to do that we have to investigate in the following energy range: $E \geq E_{L_2}$ or $E \geq E_{L_2^{'}}$. Now for simpler analysis we replace $E$ with the dimensionless energy parameter $C$ \citep{Ernst2008}, which is defined as $$ \text{for model $1$:} \; C = \frac{E_{L_2} - E}{E_{L_2}} = \frac{E_{L_4} - E}{E_{L_4}} \;\; (\because E_{L_2} = E_{L_4})$$
$$ \text{for model $2$:} \; C = \frac{E_{L_2^{'}} - E}{E_{L_2^{'}}} = \frac{E_{L_4^{'}} - E}{E_{L_4^{'}}} \;\; (\because E_{L_2^{'}} = E_{L_4^{'}}).$$ For $C>0$ orbits are unbounded and escapes through the two symmetrical channels near $L_2$. To study escape motion around $L_2$ we have to choose energy levels higher than the escape threshold. We choose our tested energy levels as $C = 0.01$ and $C = 0.1$. Energy value at $L_2$ for different values of $C$ are given in Table \ref{tab:4}. In the phase space for large sea of initial conditions we consider only those initial conditions which correspond to the central barred region, i.e. if $(x_0, y_0)$ is an initial condition in $x$ - $y$ plane then, $x_{0}^2 + y_{0}^2 \leq r_{L_2}^2$, where $r_{L_2}$ is the radial length of $L_2$. The same formalism is carried out for the point $L_2^{'}$ also.
\begin{table}[H]
\centering
\begin{tabular}{|c|c|c|}
\hline
$C$     & $E$              & $E$\\
        & (Model $1$)      & (Model $2$)\\
\hline
\hline
$0.010$ & $\;-99.81490600$ &  $-115.29682674$\\
$0.100$ & $\;-90.74082363$ &  $-104.81529704$\\
\hline
\end{tabular}
\caption{Energy levels for both the models.}
\label{tab:4}
\end{table}

\indent In our work regular and chaotic motions have been classified with the help of chaos detector MLE. For a given initial condition vector in phase space, let $\delta x(t_0)$ be the initial separation vector of two neighbouring trajectories, where $t_0$ is the initial time. Also let $\delta x(t)$ be the separation vector at time $t$. Then MLE for that designated initial condition vector is defined as,
\begin{equation}
\label{eq:5}
\text{MLE} = \lim_{t \to \infty} \lim_{|\delta x(t_0)| \to 0} \frac{1}{t} \ln \frac{|\delta x(t)|}{|\delta x(t_0)|}.
\end{equation}

\indent To follow the evolution of orbits in long time we choose our integration time as $10^2$ units, which is equivalent to $10^{10}$ years (typical age of barred galaxies). In this vast integration time, orbits starting from an initial condition will reveal their true nature (regular or chaotic). We use a set of $\tt{MATLAB}$ programmes in order to integrate the system of Eqs. (\ref{eq:3}). In order to do that we use the $\tt{ode45}$ $\tt{MATLAB}$ package with small time step $10^{-2}$. Here all the calculated values are corrected up-to eight decimal places. All the presented graphics are produced by using $\tt{MATLAB \; (version}$ - $\tt{2015a)}$ software.

\subsection{Orbital structures}
\label{sec:3.1}
\begin{figure}[H]
\centering
\includegraphics[height=8cm,width=\linewidth]{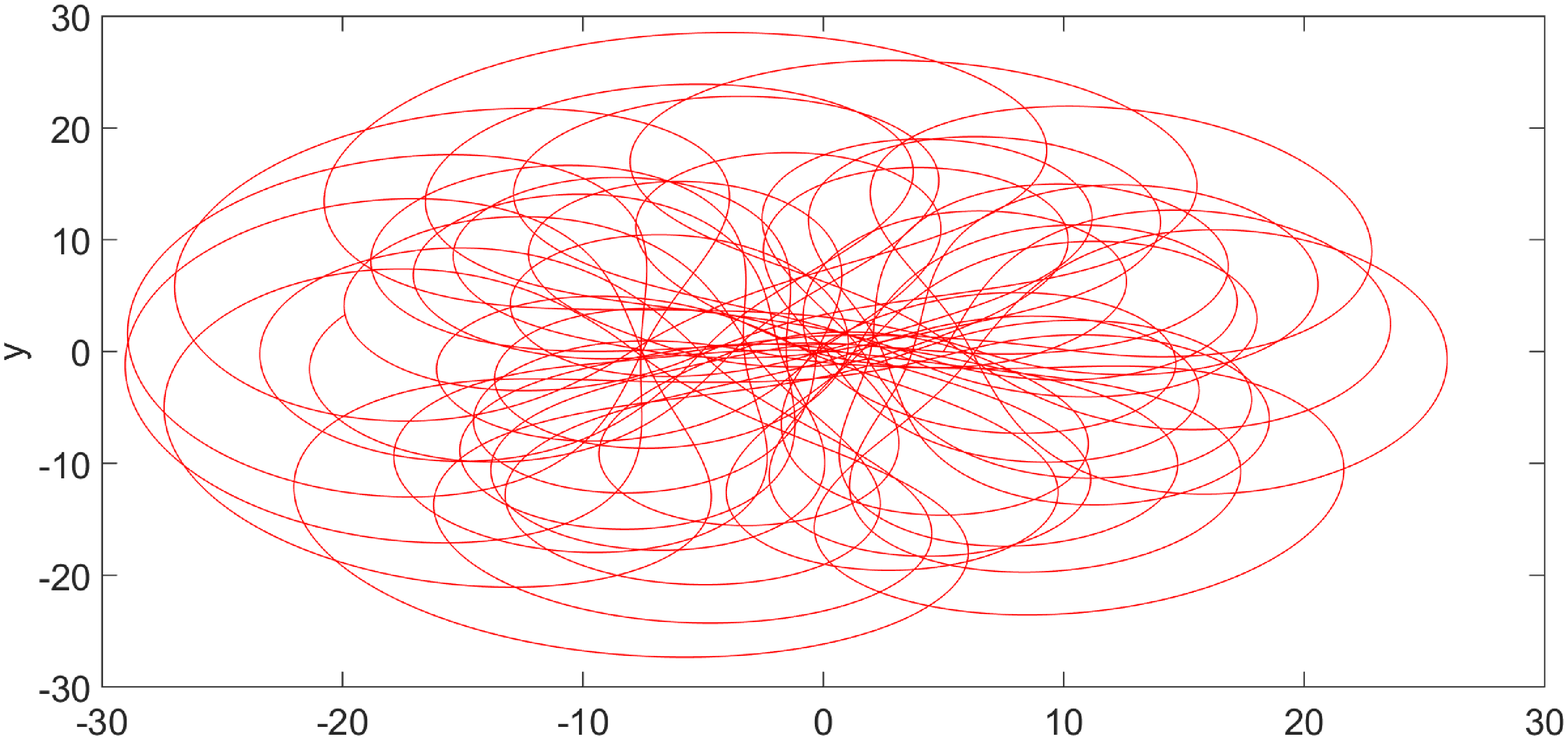}
\caption{Model $2$ -- non-escaping chaotic orbit for $C = 0.01$ with $(x_0,y_0,p_{x_0})$ $\equiv (5,0,15)$.}
\label{fig:8}
\end{figure}

\begin{figure}[H]
\centering
\includegraphics[height=8cm,width=\linewidth]{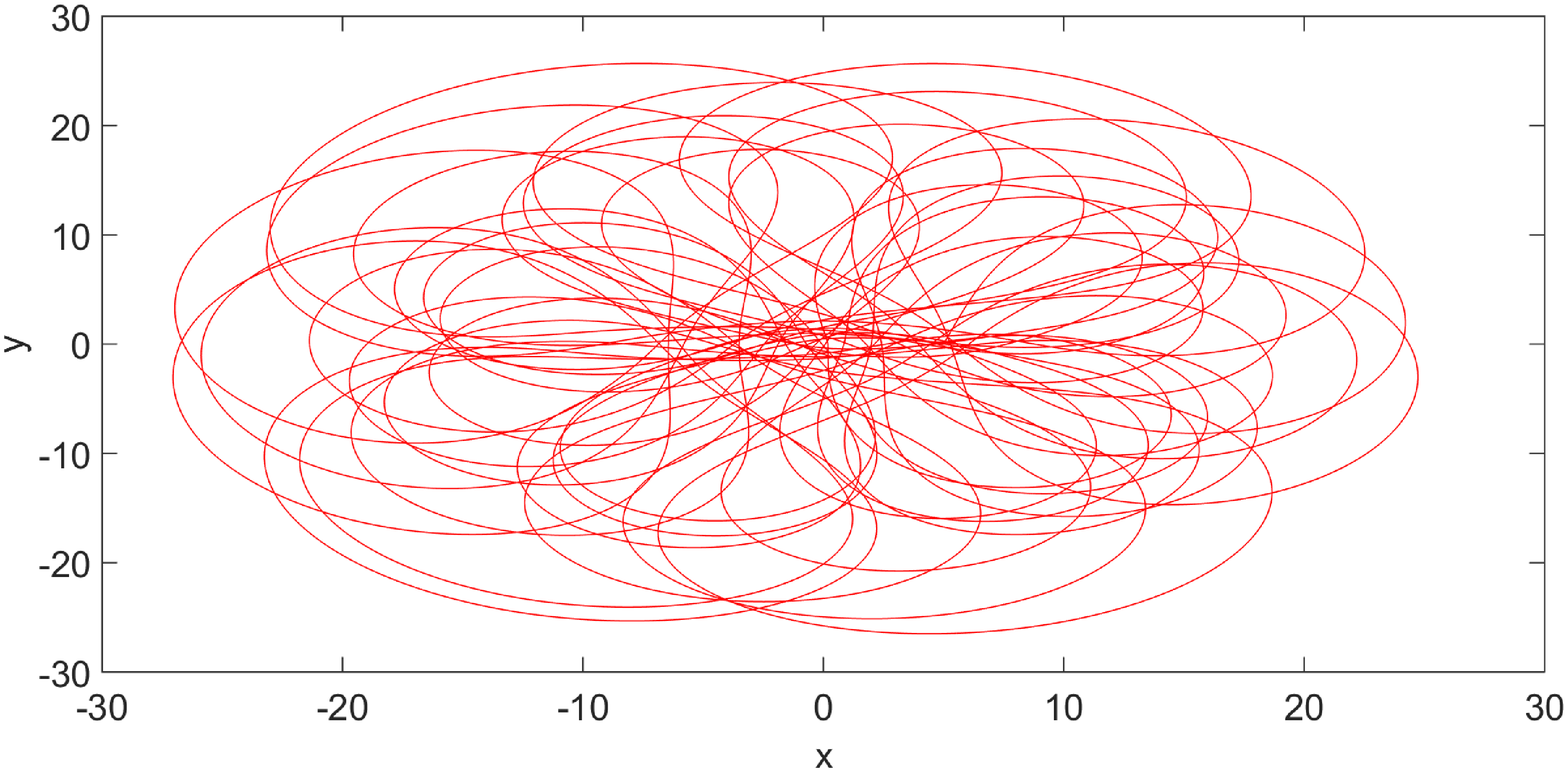}
\caption{Model $2$ -- non-escaping chaotic orbit for $C = 0.1$ with $(x_0,y_0,p_{x_0})$ $\equiv (5,0,15)$.}
\label{fig:9}
\end{figure}

\begin{figure}[H]
\centering
\includegraphics[height=8cm,width=\linewidth]{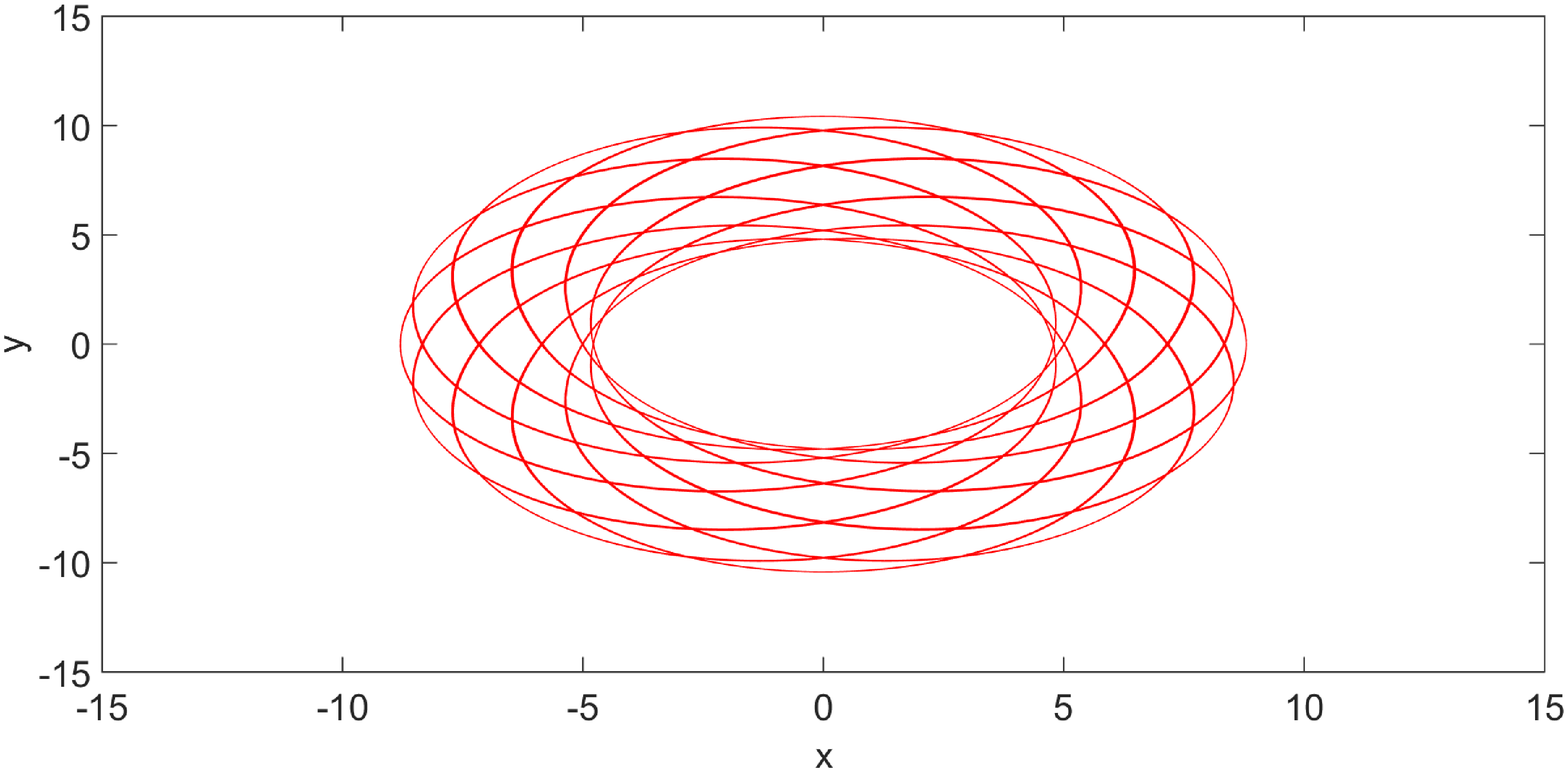}
\caption{Model $2$ -- non-escaping retrograde quasi-periodic rosette orbit for $C = 0.01$ with $(x_0,y_0,p_{x_0})$ $\equiv (-5,0,15)$.}
\label{fig:10}
\end{figure}

\begin{figure}[H]
\centering
\includegraphics[height=8cm,width=\linewidth]{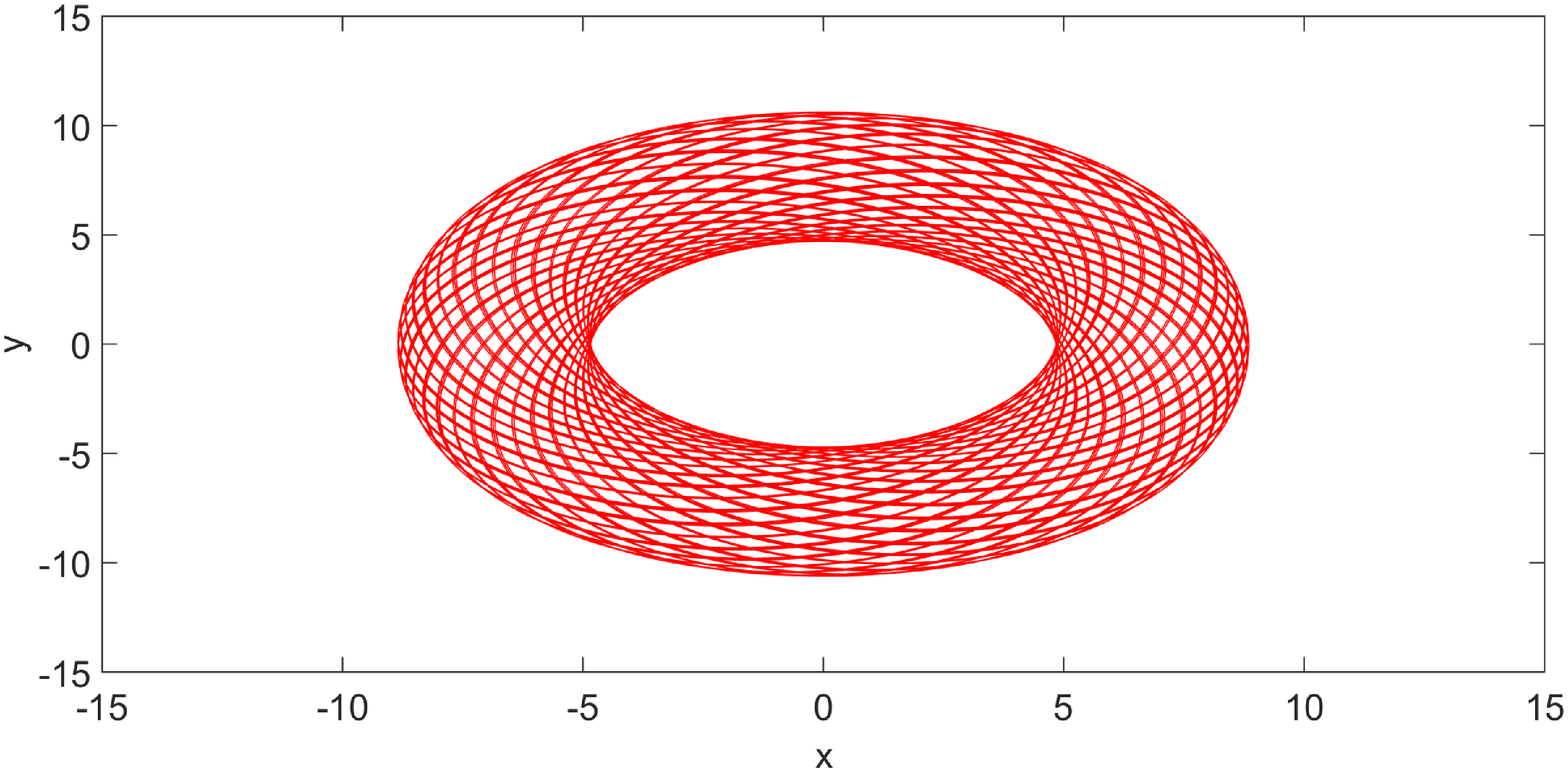}
\caption{Model $2$ -- non-escaping retrograde quasi-periodic rosette orbit for $C = 0.1$ with $(x_0,y_0,p_{x_0})$ $\equiv (-5,0,15)$.}
\label{fig:11}
\end{figure}

\noindent For studying the orbital dynamics in $x$ - $y$ plane we choose two initial conditions. One initial condition is $(x_0,y_0,p_{x_0})$ $\equiv (5,0,15)$, which help us to study the orbital dynamics in the vicinity of the Lagrangian points $L_2$ and $L_2^{'}$ respectively. Also in the vicinity of this initial condition we can figure out the properties of escape dynamics due to $L_2$ and $L_2^{'}$ at the nearest end of the bar. Similarly in order to figure out how $L_2$ and $L_2^{'}$ affects the orbital dynamics for an initial condition starting with a point away from it we choose another initial condition as $(x_0,y_0,p_{x_0})$ $\equiv (-5,0,15)$. For each of the initial condition $p_{y_0}$ is calculated from Eq. (\ref{eq:2}).
\begin{itemize}
\item Model $1$: In Figs. \ref{fig:4} and \ref{fig:5}, stellar orbits in $x$ - $y$ plane have been plotted for values $C = 0.01$ and $0.1$ respectively, with initial condition $x_0 = 5$, $y_0 = 0$ and $p_{x_0} = 15$ . In both cases we get escaping chaotic orbits. Similarly for Figs. \ref{fig:6} and \ref{fig:7} initial condition is $x_0 = -5$, $y_0 = 0$ and $p_{x_0} = 15$. Here in both cases we get non-escaping retrograde quasi-periodic rosette orbits. Value of $p_{y_0}$ for each figure is evaluated from Eq. (\ref{eq:2}) and the corresponding values of MLE are given in Table \ref{tab:5}.

\begin{table}[H]
\centering
\begin{tabular}{|c|c|c|}
\hline
Initial Condition   &$C$    &MLE\\
\hline
\hline
$(x_0,y_0,p_{x_0})$ &$0.01$ &$0.18041129$\\
$\equiv (5,0,15)$   &$0.1$  &$0.19086893$\\
\hline
$(x_0,y_0,p_{x_0})$ &$0.01$ &$0.08539470$\\
$\equiv (-5,0,15)$  &$0.1$  &$0.08948487$\\
\hline
\end{tabular}
\caption{Model $1$ -- MLE for different values of $C$.}
\label{tab:5}
\end{table}
 
\begin{figure}[H]
\centering
\includegraphics[height=8cm,width=\linewidth]{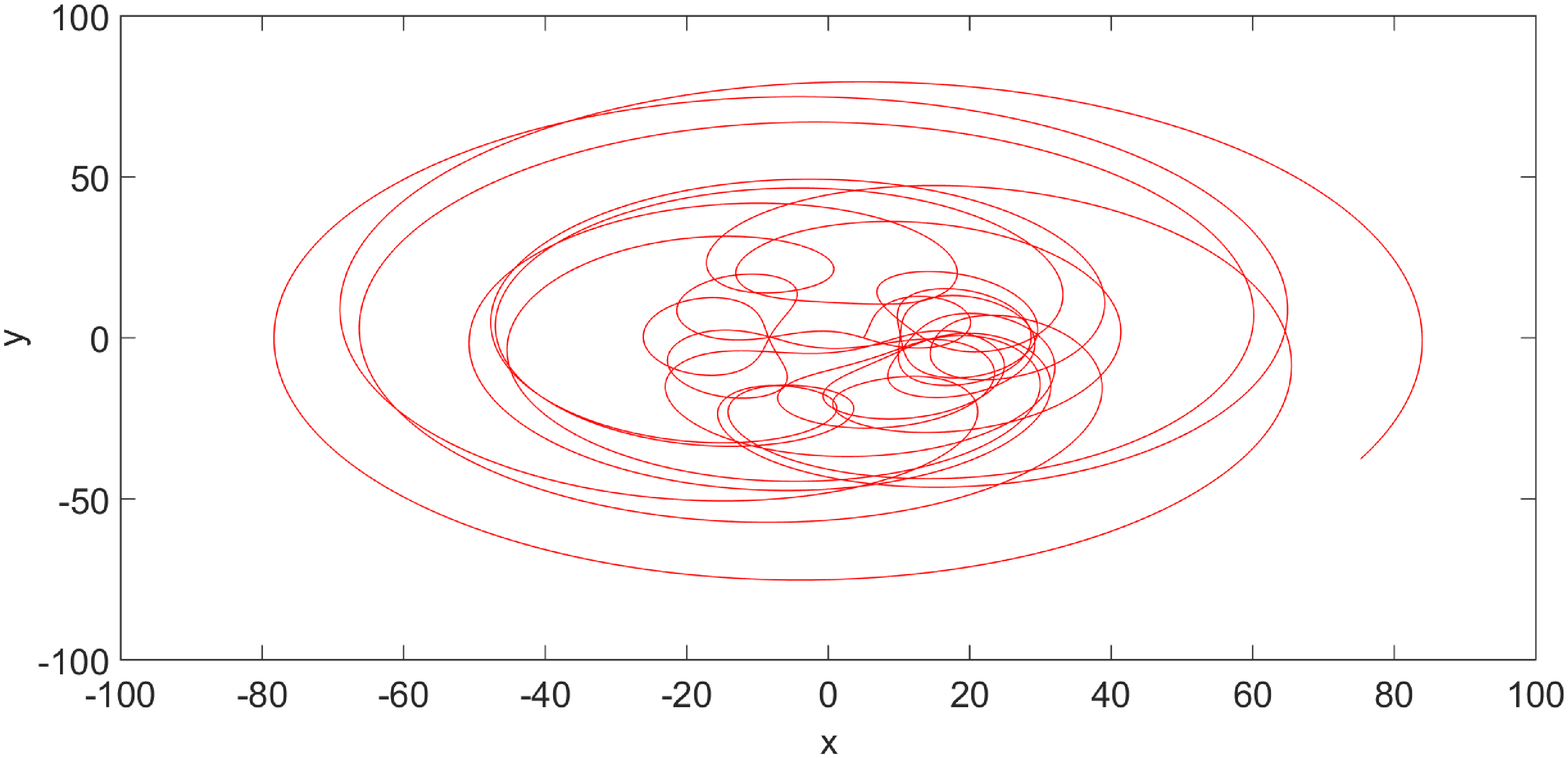}
\caption{Model $1$ -- escaping chaotic orbit for $C = 0.01$ with $(x_0,y_0,p_{x_0})$ $\equiv (5,0,15)$.}
\label{fig:4}
\end{figure}

\begin{figure}[H]
\centering
\includegraphics[height=8cm,width=\linewidth]{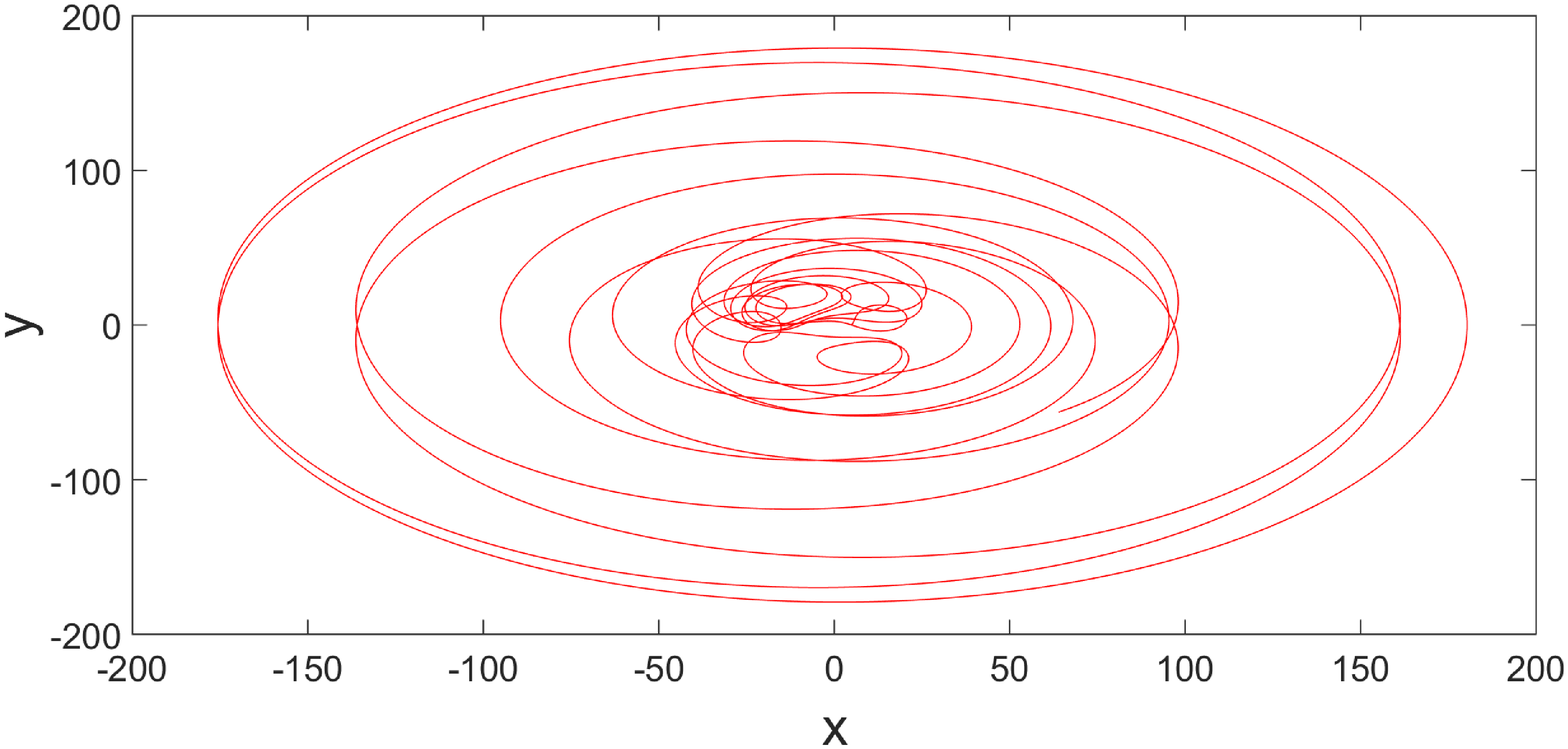}
\caption{Model $1$ -- escaping chaotic orbit for $C = 0.1$ with $(x_0,y_0,p_{x_0})$ $\equiv (5,0,15)$.}
\label{fig:5}
\end{figure}

\begin{figure}[H]
\centering
\includegraphics[height=8cm,width=\linewidth]{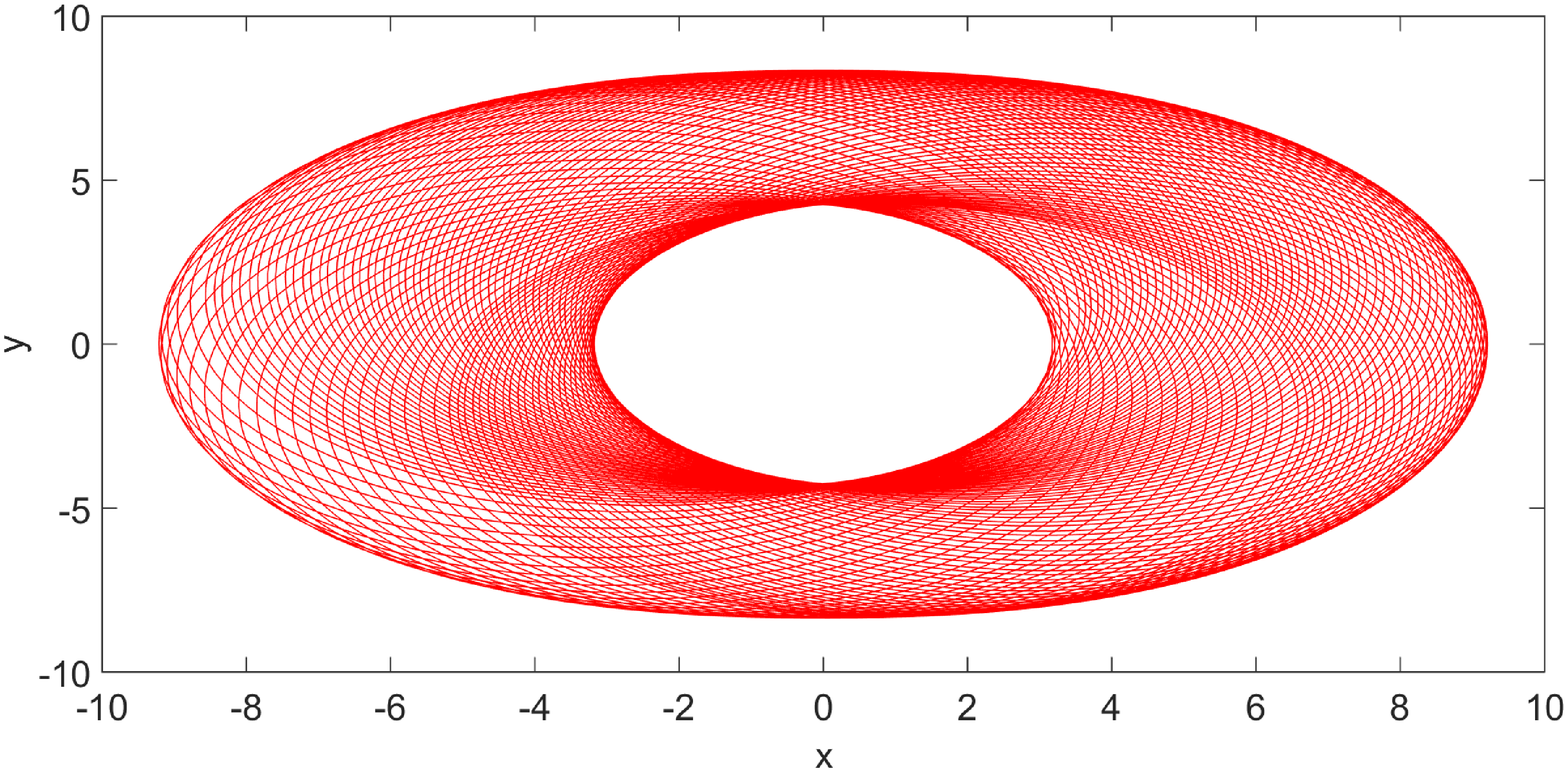}
\caption{Model $1$ -- non-escaping retrograde quasi-periodic rosette orbit for $C = 0.01$ with $(x_0,y_0,p_{x_0})$ $\equiv (-5,0,15)$.}
\label{fig:6}
\end{figure}

\begin{figure}[H]
\centering
\includegraphics[height=8cm,width=\linewidth]{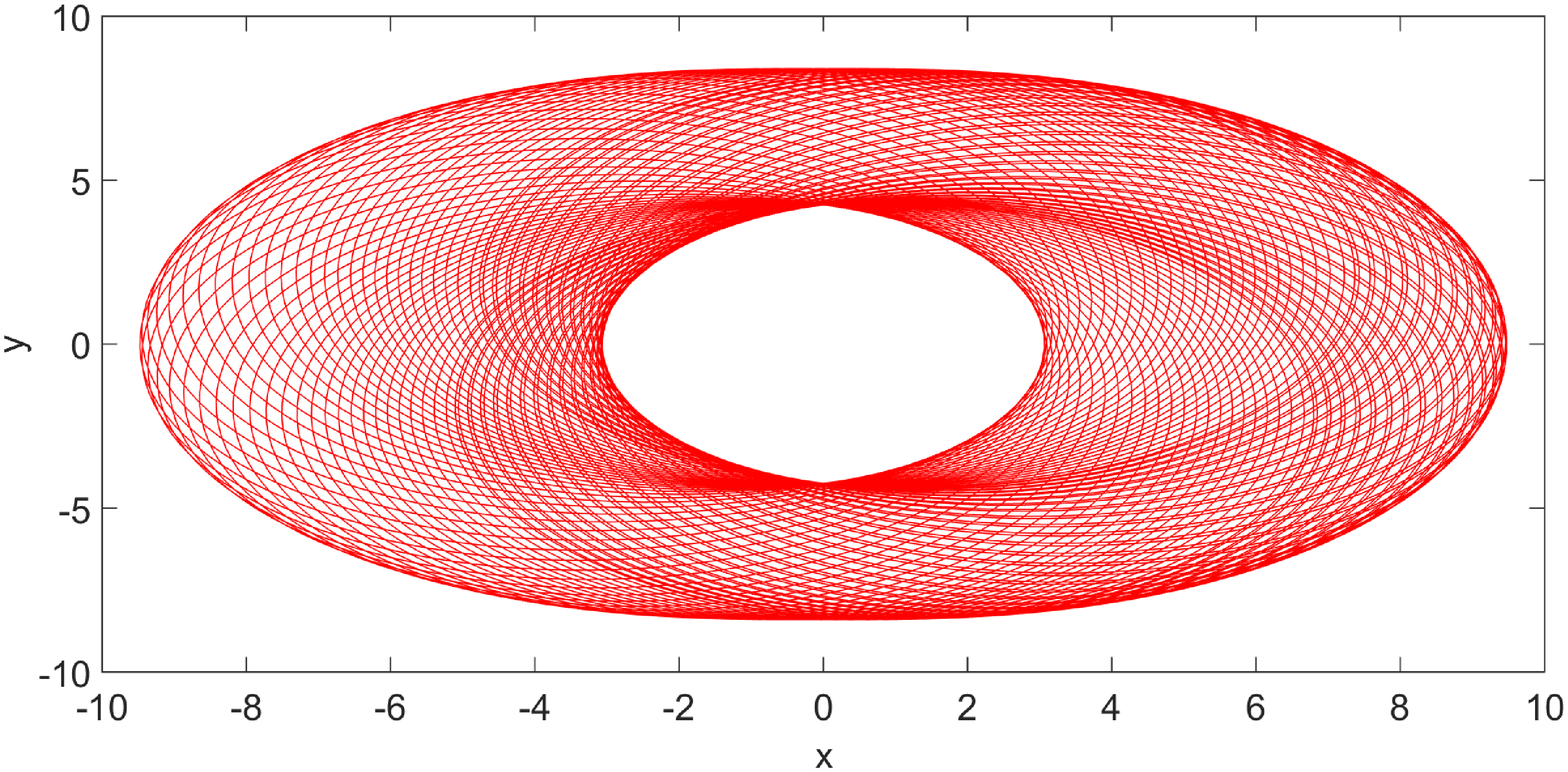}
\caption{Model $1$ -- non-escaping retrograde quasi-periodic rosette orbit for $C = 0.1$ with $(x_0,y_0,p_{x_0})$ $\equiv (-5,0,15)$.}
\label{fig:7}
\end{figure}
 
\item Model $2$: Similar as model $1$, here also in Figs. \ref{fig:8} and \ref{fig:9} stellar orbits in $x$ - $y$ plane have been plotted for values $C = 0.01$ and $0.1$ respectively with initial condition $x_0 = 5$, $y_0 = 0$ and $p_{x_0} = 15$. In both cases we get non-escaping chaotic orbits. Similarly for Figs. \ref{fig:10} and \ref{fig:11} initial condition is $x_0 = -5$, $y_0 = 0$ and $p_{x_0} = 15$. Here in both cases we get non-escaping retrograde quasi-periodic rosette orbits. Also value of $p_{y_0}$ for each figure is evaluated from Eq. (\ref{eq:2}) and corresponding values of MLE are given in Table \ref{tab:6}.

\begin{table}[H]
\centering
\begin{tabular}{|c|c|c|}
\hline
Initial Condition   &$C$    &MLE\\
\hline
\hline
$(x_0,y_0,p_{x_0})$ &$0.01$ &$0.16912615$\\
$\equiv (5,0,15)$   &$0.1$  &$0.17423968$\\
\hline
$(x_0,y_0,p_{x_0})$ &$0.01$ &$0.07609452$\\
$\equiv (-5,0,15)$  &$0.1$  &$0.07688508$\\
\hline
\end{tabular}
\caption{Model $2$ -- MLE for different values of $C$.}
\label{tab:6}
\end{table}
\end{itemize}

\subsection{Poincaré maps}
\label{sec:3.2}
\begin{figure}[H]
\centering
\includegraphics[height=8cm,width=\linewidth]{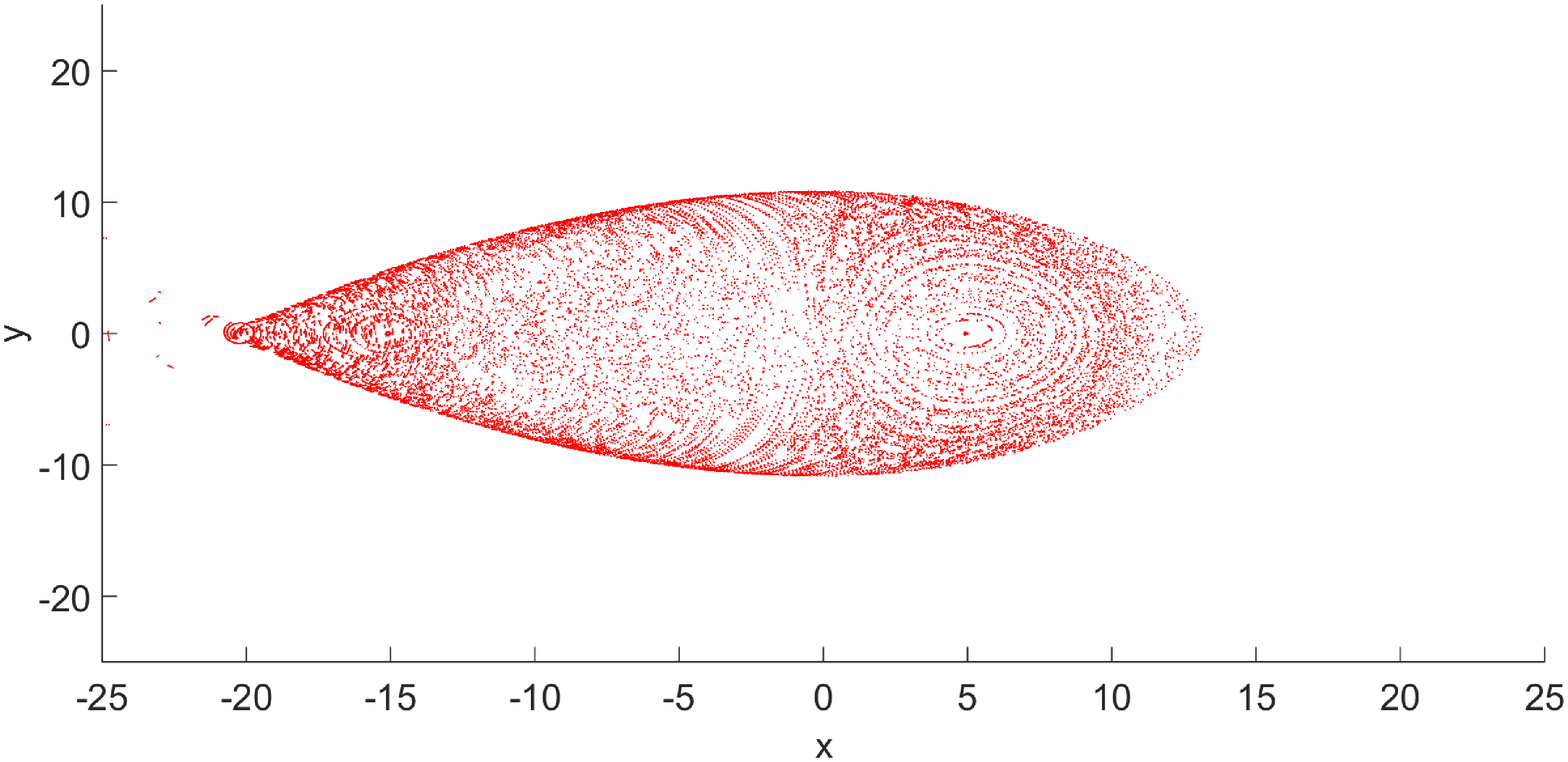}
\caption{Model $1$ -- Poincaré surface sections of $p_x = 0$ and $p_y \leq 0$ for $C = 0.01$.}
\label{fig:12}
\end{figure}

\begin{figure}[H]
\centering
\includegraphics[height=8cm,width=\linewidth]{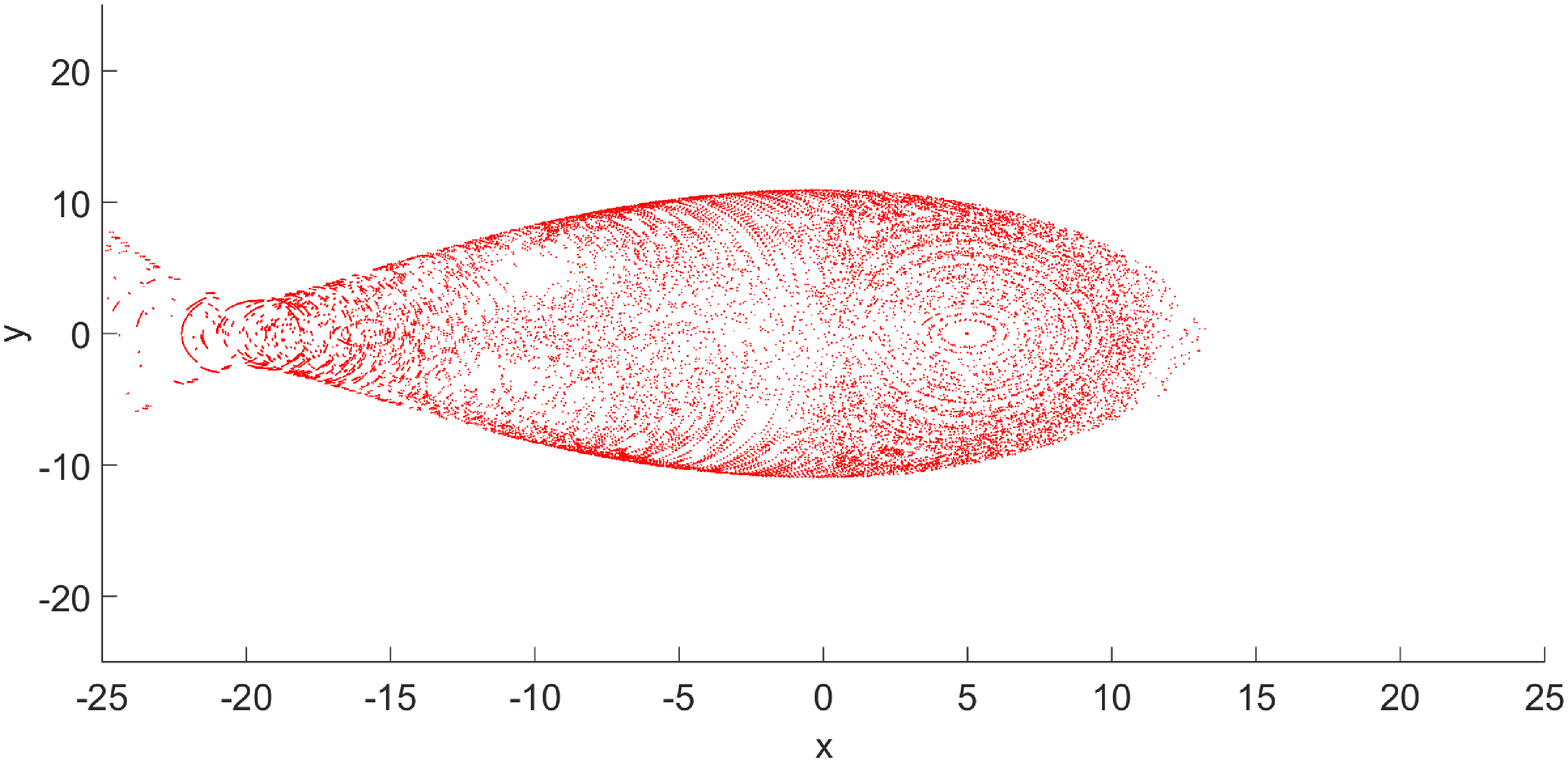}
\caption{Model $1$ -- Poincaré surface sections of $p_x = 0$ and $p_y \leq 0$ for $C = 0.1$.}
\label{fig:13}
\end{figure}

\begin{figure}[H]
\centering
\includegraphics[height=8cm,width=\linewidth]{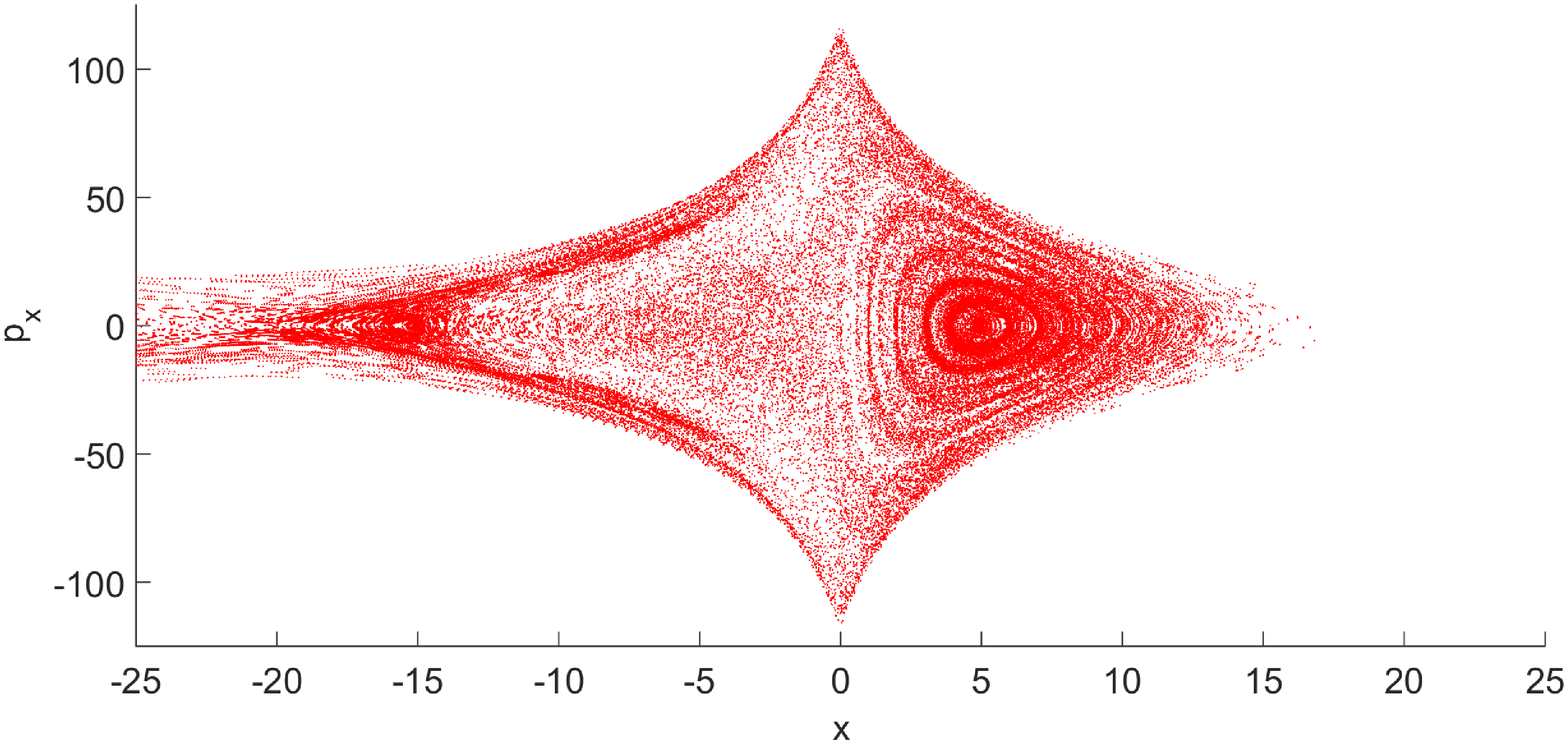}
\caption{Model $1$ -- Poincaré surface sections of $y = 0$ and $p_y \leq 0$ for $C = 0.01$.}
\label{fig:14}
\end{figure}

\begin{figure}[H]
\centering
\includegraphics[height=8cm,width=\linewidth]{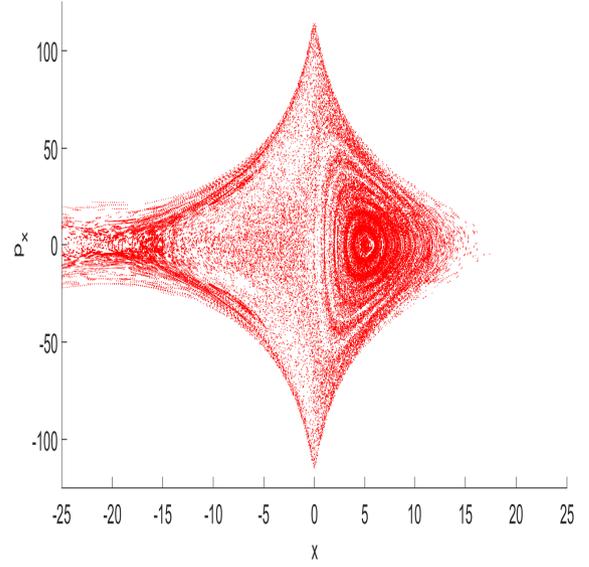}
\caption{Model $1$ -- Poincaré surface sections of $y = 0$ and $p_y \leq 0$ for $C = 0.1$.}
\label{fig:15}
\end{figure}

\begin{figure}[H]
\centering
\includegraphics[height=8cm,width=\linewidth]{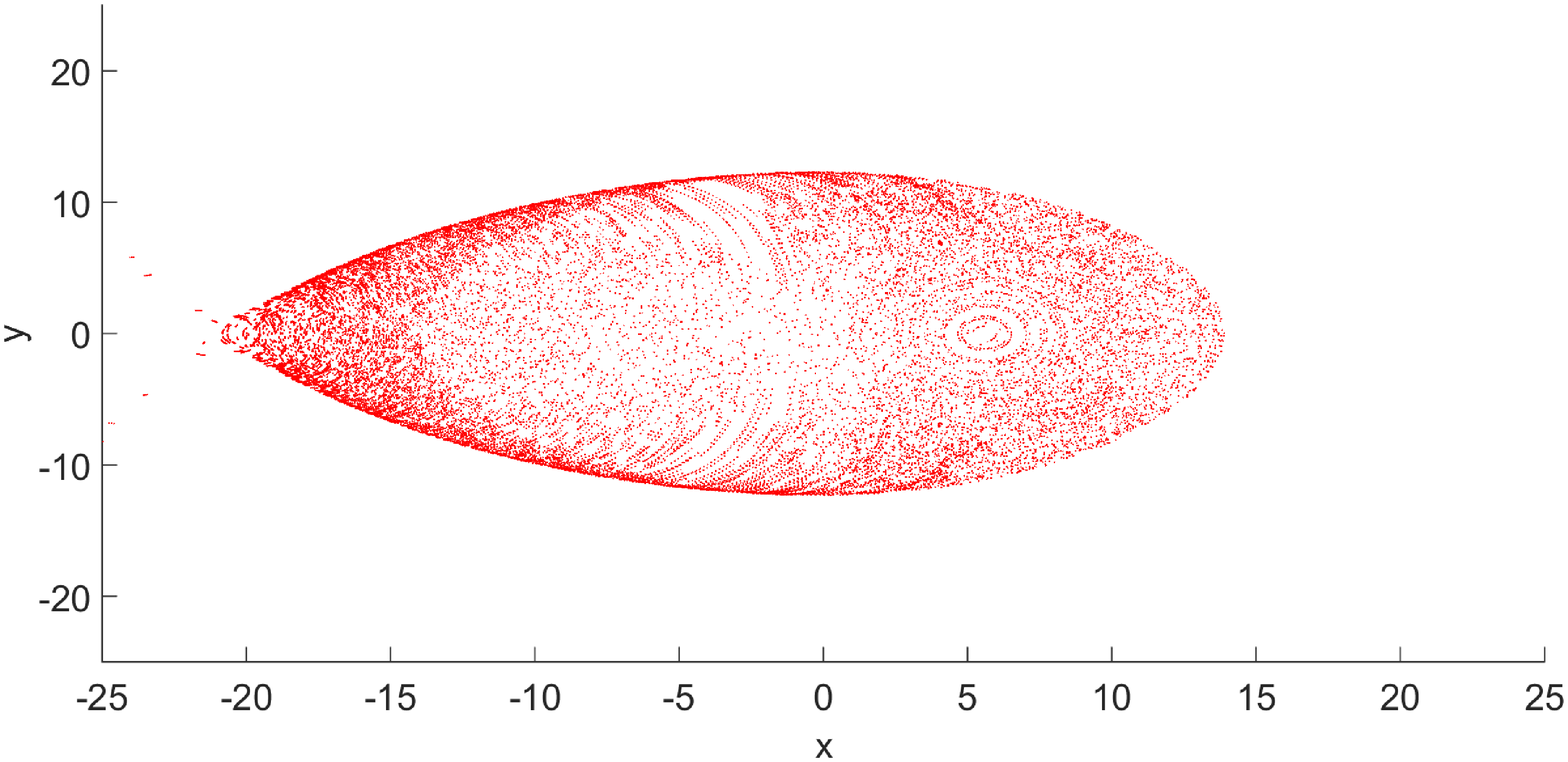}
\caption{Model $2$ -- Poincaré surface sections of $p_x = 0$ and $p_y \leq 0$ for $C = 0.01$.}
\label{fig:16}
\end{figure}

\begin{figure}[H]
\centering
\includegraphics[height=8cm,width=\linewidth]{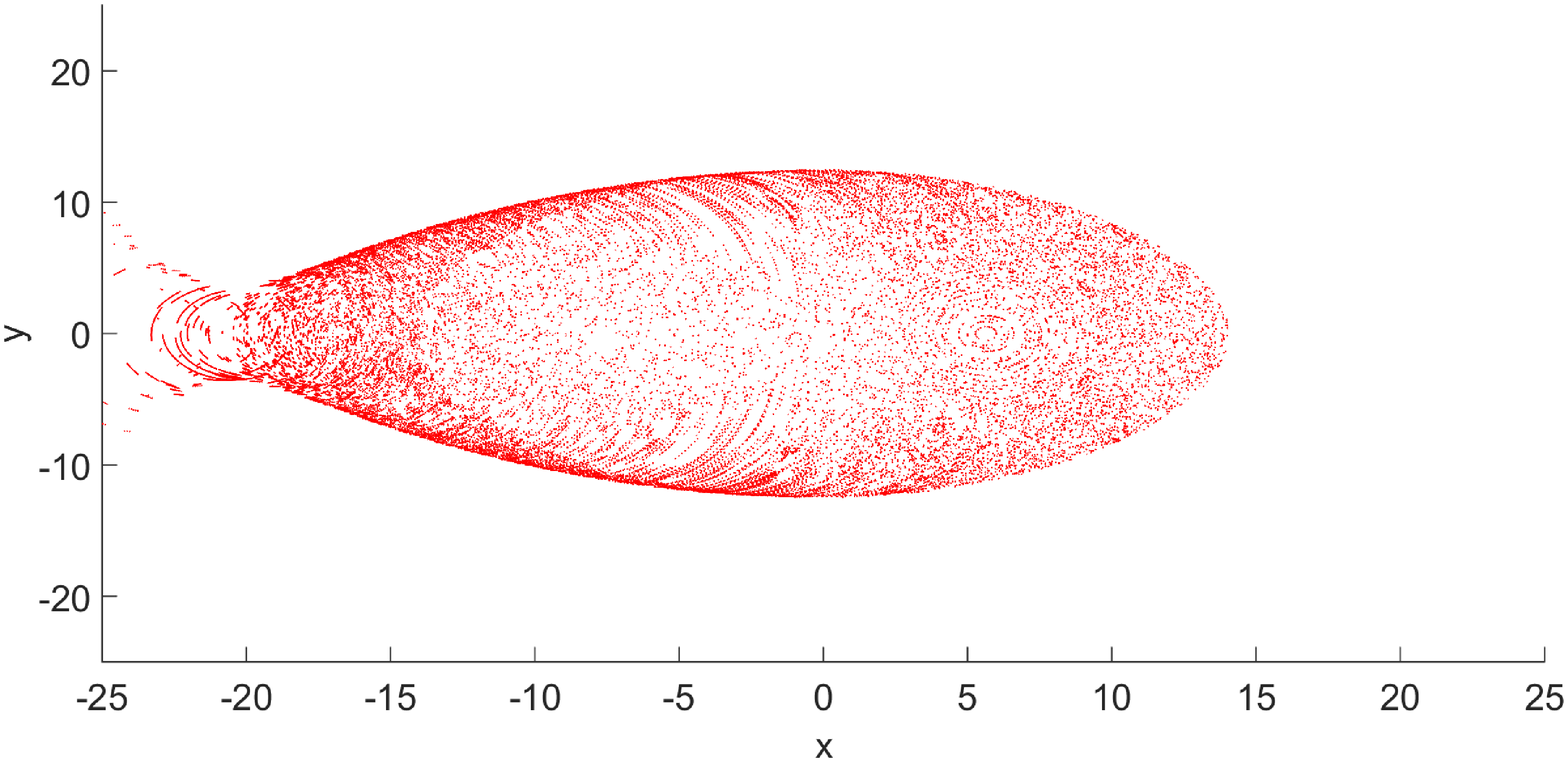}
\caption{Model $2$ -- Poincaré surface sections of $p_x = 0$ and $p_y \leq 0$ for $C = 0.1$.}
\label{fig:17}
\end{figure}

\begin{figure}[H]
\centering
\includegraphics[height=8cm,width=\linewidth]{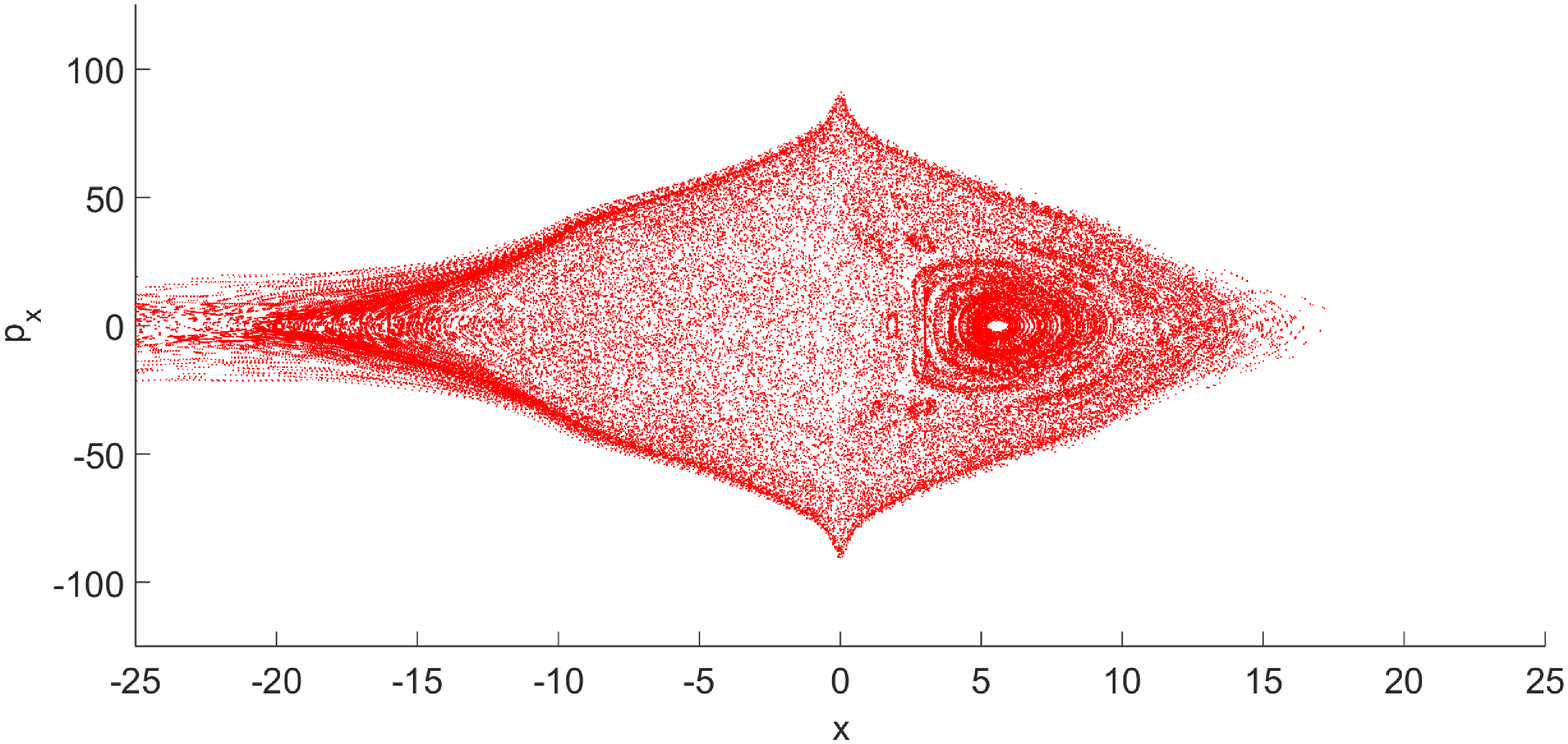}
\caption{Model $2$ -- Poincaré surface sections of $y = 0$ and $p_y \leq 0$ for $C = 0.01$.}
\label{fig:18}
\end{figure}

\begin{figure}[H]
\centering
\includegraphics[height=8cm,width=\linewidth]{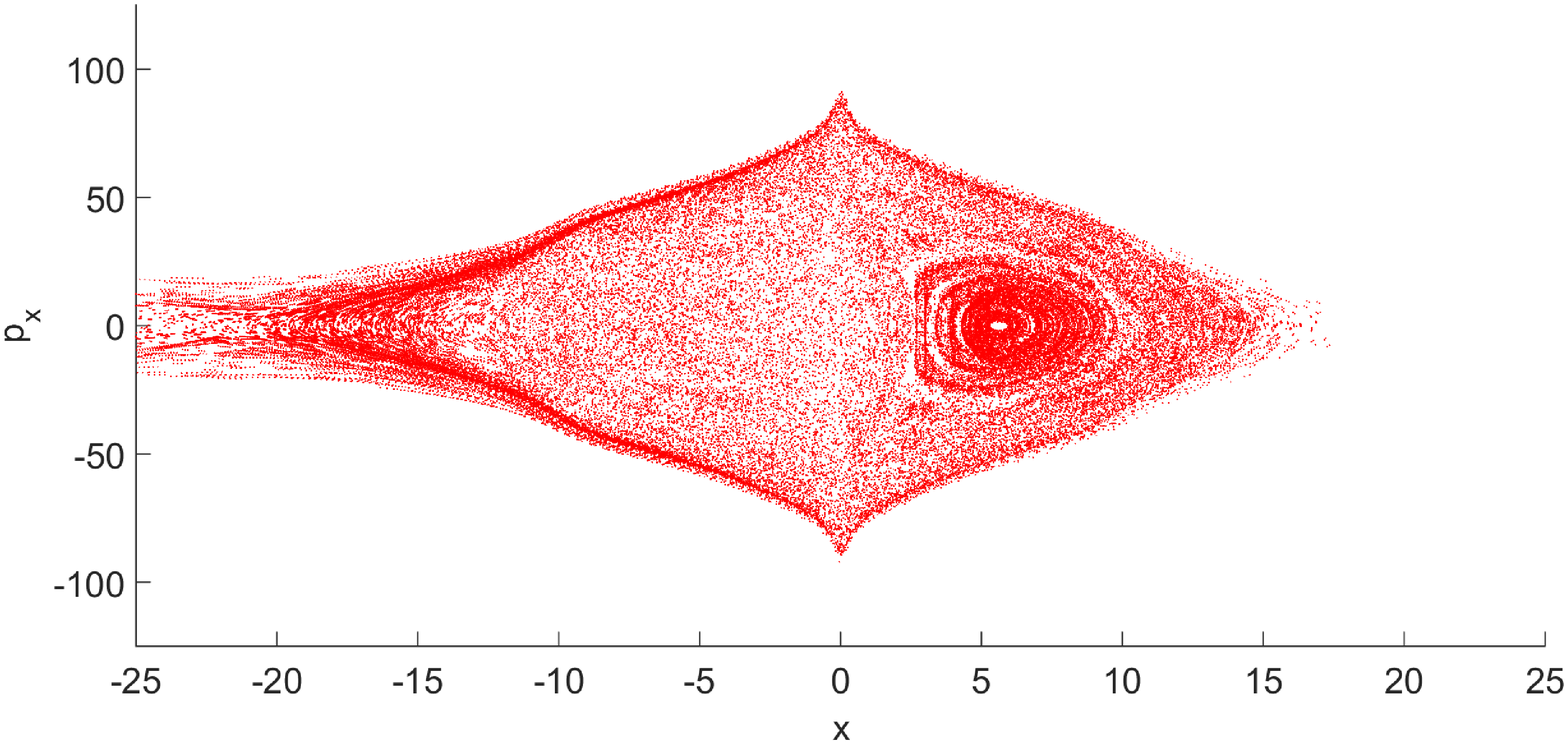}
\caption{Model $2$ -- Poincaré surface sections of $y = 0$ and $p_y \leq 0$ for $C = 0.1$.}
\label{fig:19}
\end{figure}

Poincaré maps are two-dimensional cuts of the six-dimensional hyper-surface in case of the given gravitational system. For model $1$ Poincaré surface section maps in $x$ - $y$ plane are plotted in Figs. \ref{fig:12} and \ref{fig:13} for different energy values. For this we consider a $43\times43$ grid of initial conditions with step sizes $\Delta x = 1$ kpc and $\Delta y = 1$ kpc. All the initial conditions in $x$ - $y$ plane are considered from the central barred region i.e. $x_0^2 + y_0^2 \leq r_{L_2}^2$. Initial conditions for $p_{x_0}$ and $p_{y_0}$ are $p_{x_0} = 0$ and $p_{y_0} (> 0)$, where $p_{y_0}$ is evaluated from Eq. (\ref{eq:2}). Also for Poincaré maps in $x$ - $y$ plane surface cross sections are $p_x = 0$ and $p_y \le 0$ \citep{Ernst2014}. Again for model $1$ Poincaré surface section maps in $x$ - $p_x$ plane are plotted in Figs. \ref{fig:14} and \ref{fig:15} for different energy values. For this we also consider a $43\times31$ grid of initial conditions with step size $\Delta x = 1$ kpc and $\Delta p_x = 10$ km $\text{s}^{-1}$. Here also all the initial conditions are taken from the central barred region. Initial conditions for $y_0$ and $p_{y_0}$ are $y_0 = 0$ and $p_{y_0} (> 0)$, where $p_{y_0}$ is evaluated from Eq. (\ref{eq:2}). Also for Poincaré maps in $x$ - $p_x$ plane surface cross sections are $y = 0$ and $p_y \le 0$ \citep{Ernst2014}. By using similar techniques of model $1$, Poincaré surface section maps in $x$ - $y$ plane are plotted in Figs. \ref{fig:16} and \ref{fig:17} and Poincaré surface section maps in $x$ - $p_x$ plane are plotted in Figs. \ref{fig:18} and \ref{fig:19} respectively for model $2$.

\begin{itemize}
\item Model $1$: In Fig. \ref{fig:12}, we observed that a primary stability island does exist near $(5,0)$ in $x$ - $y$ plane for energy value $C=0.01$, which has formed due to quasi-periodic motions. Again when the energy value is increased to $C=0.1$ (see Fig. \ref{fig:13}), then this stability island is gradually destroyed because of chaotic motions in that region. There is also a hint of escaping orbits in both Figs. \ref{fig:12} and \ref{fig:13} respectively. The amount of escaping orbits has increased with the increase of $C$. Similarly a primary stability island has been observed in Figs. \ref{fig:14} and \ref{fig:15} and it exists near $(5,0)$ in $x$ - $p_x$ plane. In $x$ - $p_x$ plane we also observed chaotic and escaping motions and amount of escaping orbits has been increased with increment of $C$.
\end{itemize} 

\begin{itemize}
\item Model $2$: Same as model $1$, here also in Figs. \ref{fig:16} and \ref{fig:17}, we observe that a primary stability island has formed due to quasi-periodic motions and it  exists near $(6,0)$ in $x$ - $y$ plane. As the energy value increased from $C=0.01$ to $C=0.1$, the stability island  gradually fades and motion becomes chaotic in that region. Also number of cross sectional points in $x$ - $y$ plane outside the central barred region (escaping points) are also increased with increment of $C$. Similarly in Figs. \ref{fig:18} and \ref{fig:19} a primary stability island has formed and it is exist near $(6,0)$ in $x$ - $p_x$ plane. Here also motion is chaotic and escaping cross sectional points in $x$ - $p_x$ plane are increased with increment of $C$.
\end{itemize}

\subsection{Evolution of chaos with respect to the bar parameters}
\label{sec:3.3}
Here we have discussed how the chaotic dynamics evolved over the vast integration time with respect to mass and length of the bar. In order to do that we have calculated MLE for different values of mass and length of the bar for each of the two bar potential model. Our main focus is to study the effect of chaotic dynamics in the vicinity of the Lagrangian points $L_2$ and $L_2^{'}$ respectively. That is why we restrict the study in this subsection only for the orbits starting with initial condition $(x_0,y_0,p_{x_0})$ $\equiv (5,0,15)$.
\begin{itemize}
\item Model $1$: In Table \ref{tab:7} we have showed how the chaotic dynamics vary with the bar flattening parameter ($\alpha$) and the energy parameter $C$ for model $1$. Similarly in Table \ref{tab:8} we have showed how the chaotic dynamics vary with the bar mass $M_b$ and the energy parameter $C$.

\begin{table}[H]
\centering
\begin{tabular}{|c|c|c|c|c|c|}
\hline
$C$    &$\alpha$ &MLE          &$C$   &$\alpha$ &MLE\\
\hline
\hline
$0.01$ &$1$      &$0.07183234$ &$0.1$ &$1$      &$0.07106754$\\
       &$2$      &$0.18041129$ &      &$2$      &$0.19086893$\\
       &$3$      &$0.17781788$ &      &$3$      &$0.19563488$\\
       &$4$      &$0.18365475$ &      &$4$      &$0.18118082$\\
       &$5$      &$0.18073230$ &      &$5$      &$0.18719583$\\
       &$6$      &$0.18371731$ &      &$6$      &$0.19558757$\\
       &$7$      &$0.17349070$ &      &$7$      &$0.18805772$\\
       &$8$      &$0.19519718$ &      &$8$      &$0.17684521$\\
       &$9$      &$0.18979577$ &      &$9$      &$0.17967805$\\
       &$10$     &$0.17651692$ &      &$10$     &$0.16761620$\\
\hline
\end{tabular}
\caption{Model $1$ -- MLE for different values of $\alpha$ and $C$ with $(x_0,y_0,p_{x_0})$ $\equiv (5,0,15)$.}
\label{tab:7}
\end{table}

\begin{table}[H]
\centering
\begin{tabular}{|c|c|c|c|c|c|}
\hline
$C$    &$M_\text{b}$  &MLE          &$C$    &$M_\text{b}$  &MLE\\
\hline
\hline
$0.01$ &$3100$ &$0.18501223$ &$0.1$ &$3100$ &$0.18209987$\\
       &$3200$ &$0.17796935$ &      &$3200$ &$0.19223607$\\
       &$3300$ &$0.16910068$ &      &$3300$ &$0.17153693$\\
       &$3400$ &$0.17535384$ &      &$3400$ &$0.18161073$\\
       &$3500$ &$0.18041129$ &      &$3500$ &$0.19086893$\\
       &$3600$ &$0.16146015$ &      &$3600$ &$0.18363177$\\
       &$3700$ &$0.16227338$ &      &$3700$ &$0.18316049$\\
       &$3800$ &$0.17063412$ &      &$3800$ &$0.17126018$\\
       &$3900$ &$0.17912548$ &      &$3900$ &$0.17076351$\\
       &$4000$ &$0.15472823$ &      &$4000$ &$0.19216234$\\
\hline
\end{tabular}
\caption{Model $1$ -- MLE for different values of $M_\text{b}$ and $C$ with $(x_0,y_0,p_{x_0})$ $\equiv (5,0,15)$.}
\label{tab:8}
\end{table}

\item Model $2$: In Table \ref{tab:9} we have showed how the chaotic dynamics varies with the length of semi-major axis of the bar ($a$) and the energy parameter $C$ for model $2$. Similarly in Table \ref{tab:10} we have showed how the chaotic dynamics varies with the bar mass $M_b$ and the energy parameter $C$.
\begin{table}[H]
\centering
\begin{tabular}{|c|c|c|c|c|c|}
\hline
$C$    &$a$ &MLE          &$C$   &$a$ &MLE\\
\hline
\hline
$0.01$ &$1$ &$0.07124746$ &$0.1$ &$1$ &$0.07305333$\\
       &$2$ &$0.07254412$ &      &$2$ &$0.07042724$\\
       &$3$ &$0.08711632$ &      &$3$ &$0.09603570$\\
       &$4$ &$0.08687308$ &      &$4$ &$0.07077280$\\
       &$5$ &$0.10126926$ &      &$5$ &$0.10935255$\\
       &$6$ &$0.16623252$ &      &$6$ &$0.17030541$\\
       &$7$ &$0.16912888$ &      &$7$ &$0.16864546$\\
       &$8$ &$0.17134276$ &      &$8$ &$0.17193346$\\
       &$9$ &$0.16778222$ &      &$9$ &$0.17342774$\\
       &$10$&$0.16912615$ &      &$10$&$0.17423968$\\
\hline
\end{tabular}
\caption{Model $2$ -- MLE for different values of $a$ and $C$ with $(x_0,y_0,p_{x_0})$ $\equiv (5,0,15)$.}
\label{tab:9}
\end{table}
\end{itemize}

\begin{table}[H]
\centering
\begin{tabular}{|c|c|c|c|c|c|}
\hline
$C$    &$M_\text{b}$  &MLE          &$C$    &$M_\text{b}$  &MLE\\
\hline
\hline
$0.01$ &$3100$ &$0.15980268$ &$0.1$ &$3100$ &$0.16818166$\\
	   &$3200$ &$0.17264740$ &      &$3200$ &$0.14629481$\\
	   &$3300$ &$0.17125352$ &      &$3300$ &$0.16626652$\\
	   &$3400$ &$0.17122480$ &      &$3400$ &$0.16573579$\\
	   &$3500$ &$0.16912615$ &      &$3500$ &$0.17423968$\\
	   &$3600$ &$0.16624252$ &      &$3600$ &$0.17377001$\\
	   &$3700$ &$0.17277701$ &      &$3700$ &$0.17434563$\\
	   &$3800$ &$0.16843627$ &      &$3800$ &$0.16179492$\\
	   &$3900$ &$0.16055233$ &      &$3900$ &$0.17043668$\\
	   &$4000$ &$0.17818486$ &      &$4000$ &$0.17069257$\\
\hline
\end{tabular}
\caption{Model $2$ -- MLE for different values of $M_\text{b}$ and $C$ with $(x_0,y_0,p_{x_0})$ $\equiv (5,0,15)$.}
\label{tab:10}
\end{table}

\section{Interpretations and conclusions}
\label{sec:4}
The present work describes the nature of orbits of the stars in barred spiral galaxies and the influence of bars along with the development of spiral arms as a result of escape mechanism.
	
\indent We have considered the stellar orbits in barred spiral galaxies in the presence of four components e.g. bulge, bar, disc and dark matter halo. We considered two types of bars namely (i) anharmonic bar, and (ii) Zotos bar. It is clear from Figs. \ref{fig:1} and \ref{fig:2} that the bar area of the latter one is more and it is more elongated in $x$ - direction than the first one. Also the density distribution Fig. (\ref{fig:3}) for bar in model $1$ is very high close to the centre i.e. the nature is one of cuspy type compared to model $2$. So it may be associated with strong bar. On the other hand in model $2$ the density distribution is rather flat and slightly rising in the central region. This kind of distribution may be associated with weak bar.

\indent Regarding the nature of orbits, it depends upon the initial conditions as well as bar potential. These orbits may be escaping and chaotic with high MLE for model $1$, whereas they are non escaping for model $2$ for a particular initial point (e.g. $x_0 = 5$, $y_0 = 0$, $p_{x_0} = 15$). On the other hand they are retrograde quasi-periodic with low MLE at $x_0 = -5$, $y_0 = 0$, $p_{x_0} = 15$. For the first situation it may result in spiral arms for escaping orbits and for the second case a ring may result. This is evident in S0 or ring galaxies where spiral arms are more or less absent \citep{Abadi1999, Van2009, Querejeta2015, Sil2018}.

\indent The second aspect is to study the nature of chaotic orbits of stars and gas in barred spiral galaxies. The understanding of the dynamics of these galaxies closely relates with chaotic motions of stars and gas in the central region. The presence of chaos is a manifestation of unstable orbits in these galaxies. The chaos propagates further over long period of time and influence the evolution of several structural components e.g. bar, disc and dark matter halo. Also integration of chaotic theory in the orbital and escaping stellar motions helps us to figure out true shape of these components. In case of barred spiral galaxies these growing instability relates with formation and strength of spiral arms.  

\indent Galactic bars are density waves in the disc. Spiral arms are thought to be the result of these outward density wave patterns. Observational studies in blue and near IR band of barred spiral galaxies confirm that spiral arms are continuations of the bar, which means spiral arms are outcomes of the bar driven mechanisms. While theoretical studies also suggested that there is a correlation between pattern speed of bar and that of spiral arms \citep{Elmegreen1985, Buta2009}. These density waves and stellar orbits usually have different rotational speeds but there exists some sort of corotation region inside the disc. Spiral arms emerge from the two ends of the bar through that corotation region of bar and disc. In this corotation region chaotic dynamics dominates. Theoretical studies confirm that theses chaotic orbits in the corotation region are the building blocks of the spiral arms \citep{Contopoulos1989, Kaufmann1996, Skokos2002}. Also a study by \citet{Patsis1997} have shown that these chaotic orbits are the reason behind the characteristic outer boxy isophotes of the nearly face-on bar of the barred spiral galaxy NGC 4314. Under suitable physical circumstances these spiral arms survived the chaotic dynamics in the corotation region of galactic disc. Stronger bar leads to the formation of grand design spirals. Nearly 70\% of the barred spirals in field have tightly wounded two-armed spiral structure, while only 30\% of the unbarred spirals in field have such kind of grand spiral pattern \citep{Elmegreen1982}. 

\indent Thus it is clear from the above discussion that the spiral arms might be the continuation of the escaping orbits of stars emerging from the end of bars driven by the chaos. In our case there are chaotic orbits when the energy exceeds the energy of second Lagrangian point (i.e. $E > E_{L_2}$ or $E > E_{L_2^{'}}$). We have computed MLE for various initial conditions as well as bar parameters. The following observations have been found.

\noindent (i) The orbital structures (Figs. \ref{fig:4} - \ref{fig:7}) for model $1$, starting at initial conditions $(x_0,y_0,p_{x_0})$ $\equiv (5,0,15)$ and $(x_0,y_0,p_{x_0})$ $\equiv (-5,0,15)$, encourage stellar orbits escaping from the ends of the bar (i.e. Lagrangian points $L_2$ and $L_4$). Also the amount of chaos increases with the energy. Similarly from Poincaré surface section maps (Figs. \ref{fig:12} - \ref{fig:15}) the similar results have been found.

\noindent (ii) For model $1$, MLE increases with the flattening parameter ($\alpha$) up to a threshold length and again slowly decreases (viz. Table \ref{tab:7}). 

\noindent (iii) The threshold length decreases with increasing value of $C$. This implies when the escape energy of $L_2$ is high ($C \sim 0.1$), escape of stars are even encouraged at smaller bar length, e.g. for $C = 0.01$, the threshold value of $\alpha = 8$ whereas for $C = 0.1$, $\alpha = 6$ (viz. Table \ref{tab:7}). 

\noindent (iv) From Table \ref{tab:8} it is clear that MLE more or less decreases with bar mass. So, heavier (or strong) bars generally oppose escape mechanism. Thus formation of spiral arms for smaller values of escape energy are not favourable but as the energy value increases (may be due to central explosions, shocks etc.) i.e. if $C \sim 0.1$ the MLE increases with mass of the bar. So in the presence of  massive strong bar  the formation of spiral arms are more likely in those barred spiral galaxies where violence activities are occurring in the central region. We all know, giant spirals harbour Super Massive Black Holes (hereafter SMBH) at their nuclear regions where violent activities are going on \citep{Basu1989, Melia2001, Kaviraj2015, Mondal2019, Kim2020}. So many giant spirals containing SMBH shows grand design spiral arms \citep{Ann2004, Pastorini2007, Seigar2008, Berrier2013, Davis2017}. Such grand design spiral arms had been found in M83, NGC 628, NGC 5457, NGC 6946 \citep{Ced2013, Frick2016, Wez2016} etc. Hence bar potential used in model $1$ favours the formation of grand design spirals when SMBH is present at the core and the spiral arms emerge from the ends of the bar. 

\noindent (v) The orbital structures (Figs. \ref{fig:8} - \ref{fig:11}) for model $2$, starting with initial conditions $(x_0,y_0,p_{x_0})$ $\equiv (5,0,15)$ and $(x_0,y_0,p_{x_0})$ $\equiv (-5,0,15)$, do not encourage stellar orbits escaping from the ends of the bar (i.e. Lagrangian points $L_2^{'}$ and $L_4^{'}$) rather the chaotic behaviour remained encapsulated inside the central region. Also the amount of chaos increases with the energy value. Poincaré surface section maps (Figs. \ref{fig:16} - \ref{fig:19}) also confirm the same phenomena. 

\noindent (vi) The MLE increases with semi-major axis of the bar ($a$) (viz. Table \ref{tab:9}) up to a threshold value ($a = 6$) and does not vary much with further flattening and it is independent of energy levels. So, for this type of bar an optimal bar length encourages escaping orbits and formation of spiral arms.

\noindent (vii) Weaker bars helps escape mechanism but the increase of chaos with bar mass is very slow (viz. Table \ref{tab:10}). This might be the reasons why spiral arms are not very prominent in many barred galaxies and as a result the formation of S0 or ring galaxies \citep{Regan2003, Byrd2006, Athanassoula2009, Baba2013, Proshina2019}. Observational evidence of such type of ring structures had been found in NGC 1326 by \citet{Buta1995}. Similarly studies by \citet{Sakamoto1999, Sakamoto2000} had also identified such ring type of structures in NGC 5005. Hence bar potential used in model $2$ favours the formation of ring type of structures or less prominent spiral arms which emerge from the ends of the bar.

\indent Thus from the above discussions we have come to the following conclusions.
\begin{itemize}
\item Barred galaxies with massive and stronger bar potential (viz. model $1$) may lead to the formation of grand design spirals only when there are some kind of violent activities going inside their central region. SMBHs may be one of the reasons for that kind of violent activities. Again galaxies with stronger or massive stronger bars but without central SMBHs may lead to the formation of less prominent spiral arms.    
\item On the contrary barred galaxies with weaker bar potential (viz. model $2$) may lead to the formation of ring type of structures or less prominent spiral arms around their central region.
\end{itemize}

\indent In a subsequent work we will discuss how different types dark matter halo profiles affect the orbital and escape dynamics in the central region and conceptualise whether they have any role behind further substructure formations or not.\\

\section*{Acknowledgements}
The author DM thanks the University Grants Commission of India for providing Junior Research Fellowship (ID - 1263/(CSIRNETJUNE2019)), under which the work has been done. 

\section*{Data availability}
\indent Both the authors confirm that the analysed data supporting the findings of this study are available within the article. 












\bsp	
\label{lastpage}
\end{document}